\documentclass[aps, prd, preprintnumbers, showpacs, amsfonts, nofootinbib, floatfix]{revtex4}      
\usepackage{graphicx,epsfig}

\usepackage[dvips]{color}

\parskip \baselineskip

\def\gsim{\lower0.5ex\hbox{$\:\buildrel >\over\sim\:$}}
\def\lsim{\lower0.5ex\hbox{$\:\buildrel <\over\sim\:$}}
\begin{document}
\preprint{CUMQ/HEP 149}
%
%
\title{\Large  The Casimir Force in Randall Sundrum Models with $q+1$ dimensions}
\author{Mariana Frank$^a$}
\email[]{mfrank@alcor.concordia.ca}
\affiliation{$^a$Department of Physics, Concordia University, 7141 Sherbrooke St. 
 West, Montreal, Quebec, CANADA H4B 1R6}
 \affiliation{$^b$Department of Mathematics and Statistics, University of Prince Edward Island, 550 University Avenue, Charlottetown, PE, CANADA C1A 4P3}

\author{Nasser Saad$^{b}$}
\email[]{nsaad@upei.ca}
\affiliation{$^a$Department of Physics, Concordia University, 7141 Sherbrooke St. 
 West, Montreal, Quebec, CANADA H4B 1R6}

\affiliation{$^b$Department of Mathematics and Statistics, University of Prince Edward Island, 550 University Avenue, Charlottetown, PE, CANADA C1A 4P3}

\author{Ismail Turan$^a$}
\email[]{ituran@physics.concordia.ca}
\affiliation{$^a$Department of Physics, Concordia University, 7141 Sherbrooke St.
 West, Montreal, Quebec, CANADA H4B 1R6}
\affiliation{$^b$Department of Mathematics and Statistics, University of Prince Edward Island, 550 University Avenue, Charlottetown, PE, CANADA C1A 4P3}

\date{\today}

\begin{abstract}
We evaluate the Casimir force between two parallel plates in Randall Sundrum (RS) scenarios extended by $q$ compact dimensions. After giving exact expressions for one extra compact dimension (6D RS model), we generalize to an arbitrary number of compact dimensions. We present the complete calculation for both two brane scenario (RSI model) and one brane scenario (RSII model) using the method of summing over the modes. We investigate the effects of extra dimensions on the magnitude and sign of the force, and comment on limits for the size and number of the extra dimensions. 

\end{abstract}
\maketitle
\section{Introduction}\label{sec:intro}
Theories with extra space dimensions have a long history in physics, at the beginning being motivated by the attempt to unify the fundamental interactions, as in the Kaluza-Klein theories \cite{Kaluza:1921tu}.
 Recently, they have enjoyed a resurgence with the development of models  with warped extra dimensions \cite{Randall:1999vf}, which  provide an explanation of the large gap between the Planck and the electroweak scale (the hierarchy problem) and the cosmological constant.
The possibility of the existence of extra dimensions, as well as their geometry and size influence the structure of the vacuum, in particular the evaluation of the vacuum zero-point energy, known as the Casimir effect  \cite{Casimir:dh}. The research in this area is motivated in two directions. First,  developments in the fundamental area of the structure of the vacuum quantum field theories have been extensively explored, with a view to understand the implications of extra dimensions  \cite{Ponton:2001hq}. Second, several measurements of the attractive force between parallel plates (and other geometries) have firmly established the existence of quantum fluctuations. The level of precision reached by these experiments may be sufficient to test models with different geometries \cite{Bordag}. Casimir forces in space-time with $D>4$ Euclidean dimensions were first calculated systematically by Ambj{\o}rn and Wolfram \cite{Ambjorn:1981xv}. Previous studies have analyzed cosmological aspects of the vacuum, such as the cosmological constant as a manifestation of the Casimir energy during the primordial cosmic inflation \cite{Peloso:2003nv}. The Casimir energy has been investigated in the context of string theories \cite{Fabinger:2000jd}, and even in Randall Sundrum models, as  means of stabilizing the radion \cite{Garriga:2002vf}. The dynamical Casimir effect has also been discussed in warped braneworlds \cite{Durrer:2007ww}. The Casimir force for parallel plate geometry has been calculated in UED \cite{Poppenhaeger:2003es,Pascoal:2007uh, Cheng:2006pe} and in various other frameworks \cite{Nam:2000cv}.  

The presence of the extra dimensions modifies the dispersion relation $\displaystyle \frac{\omega^2}{c^2}=k^2_{4D}+ \Delta k^2_{\rm x D}$ where $k_{4D}$ is the usual 4 dimensional wave-vector and $\Delta k^2_{\rm x D}$ would depend on specifics of the model used. By applying strict experimental constraints to the resulting energy density or force, one can restrict the size or curvature of the new space dimensions. 
In a previous work \cite{Frank:2007jb}, two of us analyzed the effects of one extra dimension on the Casimir force and energy in 5D dimensional  warped space time of models RSI and RSII. In this work we extend our 5D calculation to Randall-Sundrum models extended by one, and then an arbitrary number of  extra compact dimensions\footnote{Additional dimensions in 5D RS models have to be compact. Otherwise one could not recover Einstein's equations.}. Such higher dimensional scenarios \cite{Gherghetta:2000qi}  are interesting because they allow for localization of gauge and matter fields.  For the extra spatial dimensions a common choice is to work in polar coordinates ($R_{\theta}, \theta$), where $0 \le R_{\theta} < \infty$ and $0 \le \theta < 2\pi$.
We expand and generalize the calculation  of the Casimir force in the Randall-Sundrum model with two branes (RSI) and and one brane (RSII) between two parallel plates with arbitrary numbers of compact dimensions. We use as before the summation of the modes method. As far as we know, there are no alternative investigations of the qD RSI model in the literature. For the one-brane case (RSII) there exists an alternative evaluation using  the Green function formalism \cite{Linares:2008am}. We compare our results with theirs and comment on any discrepancies.

Our paper is organized as follows. In the next section, we introduce some general features of the model in 6 dimensions and derive the dispersion relation.  In section III we present our analysis of the Casimir force in RSI models. We start with the 6 dimensional model (one extra compact dimension) and then generalize to an arbitrary number. We do the same for the RSII model in section IV. The numerical and graphical analysis of our results is included in section V, where we try to restrict the size and number of extra dimensions. We compare our results with others, summarize and conclude in section VI.

\section{Randall Sundrum Model with extra compact dimensions}
\label{sec2}
\noindent The metric of 5D space-time Randall-Sundrum (RS) model with 4D Poincar{\'e} invariance is
\begin{equation}
 ds^2_{5D}=e^{-2 \mathtt{k} |y|}\eta_{\mu\nu}dx^\mu dx^\nu-dy^2\,,
\end{equation}
where $\mathtt{k}^{-1}$ is the radius of 5D Anti-de Sitter $AdS_5$ space,  $y$ is the spatial orbifolded extra dimension and $\eta_{\mu\nu}$ is the usual 4D metric with the convention $diag\,(+,-,-,-)$. 

The generalization to more than one extra spatial dimension can also be done, and as mentioned earlier, these dimensions must be compact. Localization of graviton on the brane motivates us to consider additional dimensions both compact and warped \cite{Rubakov:2001kp}. The metric in generalized $(3+q+2)$ dimensions (where $3+1$ are the usual space-time dimensions, there is one  extra $y$ spatial dimension, and the extra $q$ are compact and warped dimensions) is \cite{Gherghetta:2000qi}
\begin{equation}
 ds^2_{5D+n}=e^{-2\mathtt{k}|y|}\left(\eta_{\mu\nu}dx^\mu dx^\nu-\sum_{j=1}^{q}R_j^2d\theta_j^2\right)-dy^2\,,
\label{eqn:ndim}
\end{equation}
where $y$ is non-compact and $\theta_j$'s are compact extra coordinates running from $0$ to $2\pi R_j$. This metric can be a solution to Einstein equations in $(5+q)$ dimensions with negative bulk cosmological constant \cite{Ambjorn:1981xv,Gherghetta:2000qi}. In the  6D case, (\ref{eqn:ndim}) for $n=1$ becomes
\begin{equation}
 ds^2_{6D}=e^{-2\mathtt{k}|y|}\left(\eta_{\mu\nu}dx^\mu dx^\nu-R^2d\theta^2\right)-dy^2\,,
\label{eqn:6dim}
\end{equation}
where we simply use $R$ and $\theta$ for the extra compact warped dimension. Here the 6D metric $g_{MN}$ is
\begin{equation}
ds^2_{6D}=g_{MN}dx^Mdx^N\,,
\end{equation}
such that $g_{MN}=e^{-2\mathtt{k} |y|}\,diag(\eta_{\mu\nu},-1,-e^{2\mathtt{k} |y|})$\footnote{Note that $g^{MN}=e^{2\mathtt{k} |y|}\,diag(\eta^{\mu\nu},-1,-e^{-2\mathtt{k} |y|})$.}. The Greek indices $\mu,\nu$ run over $0$ to $3$ while Roman indices $M,N$ run over $0$ to $5$. 

As in our 5D calculation \cite{Frank:2007jb} , we use the scalar field analogy to represent the photon in 6D.  The field equation for a scalar field $\phi$ in 6D with mass $m_6$ can be obtained from the Lagrange density ${\cal L}(\phi,\partial_M \phi)$,
\begin{equation}
{\cal L}(\phi,\partial_M \phi)=\sqrt{-g}\left(\frac{1}{2}g^{MN}\partial_M\phi\partial_N\phi-\frac{1}{2}m_6\phi^2\right),
\end{equation}
where $g$ is the determinant of the metric $g_{MN}$ which is $y$-dependent. (This has discussed previously in \cite{Linares:2007yz} and we include it here for completeness.) The field equation for $\phi$ becomes
\begin{equation}
\left(e^{2\mathtt{k} |y|}\eta^{\mu\nu}\partial_\mu \partial_\nu-R^{-2}e^{2\mathtt{k} |y|}\partial_\theta^2-\sqrt{-g}^{-1}\partial_y\left(\sqrt{-g}\partial_y\right)+m_6^2\right)\phi=0.
\label{eqn:EOM}
\end{equation}
By using the separation of variables for $\phi(x^\mu,\theta,y)=X(x^\mu)Y(y)\Theta(\theta)$, one gets for $y>0$
\begin{eqnarray}
&&\frac{d^2Y(y)}{dy^2}-5\mathtt{k}\,\frac{dY(y)}{dy}+{m}^2e^{2\mathtt{k} y}Y(y)-m_6^2Y(y)=0\,,\label{eqn:y-eqn}\\
&&\eta^{\mu\nu}\frac{d^2X(x)}{dx^\mu dx^\nu}+m_{eff}^2X(x)=0\,,\label{eqn:4Deqn}\\
&&\Theta_n(\theta)=\frac{1}{\sqrt{2\pi R}}e^{in\theta}, n=0,1,2...
\end{eqnarray}
where $m_{eff}^2\equiv {m}^2+n^2/R^2$ is the effective mass in 4D.
Since we use the scalar field-photon analogy, we consider the massless scalar field\footnote{There are some issues with the massive case in 6D for RSII since there are no zero modes. See \cite{Linares:2007yz} for details.}.

Equation (2.7) can be solved directly using the master formula for the equation of the form \cite{Polyanin and Zaitsev}: 
\begin{equation}\label{m3}
y''+ay'+(be^{\lambda x}+c)y=0.
\end{equation}
The solution for (2.7) with $m_6\ne0$, with an appropriate choice of parameters, becomes:
\begin{eqnarray}\label{m6}
Y&=&e^{5\mathtt{k} y/2}[C_1 J_\nu({1\over \mathtt{k}}{{m}}e^{\mathtt{k} 
y})+C_2Y_\nu({1\over\mathtt{k}}{m}e^{\mathtt{k} y})]
\end{eqnarray}
where $J_{\nu}(z)$ and $Y_{\nu}(z)$ are 
the Bessel functions of the first and second kind and 
$\nu={1\over 2\mathtt{k}}\sqrt{25\mathtt{k}^2+4m_6^2}$.  For the massless case $\nu=5/2$.

We apply the Neumann boundary conditions $\partial_y Y(0)=0=\partial_y Y(\pi 
R_0)$, to obtain the masses ${m}$ as solutions to the 
transcendental equation
\begin{equation}\label{m27}
  J_{3\over 2}({{m}\over \mathtt{k}})Y_{3\over 2}({{m}\over 
\mathtt{k}}e^{\pi \mathtt{k}R_0})- Y_{3\over 2}({{m}\over \mathtt{k}})  J_{3\over 
2}({{m}\over \mathtt{k}}e^{\pi \mathtt{k}R_0}) = 0
\end{equation}
Expressing the  Bessel functions  $J_{n+{1\over 2}}(x)$ in 
terms of elementary functions we get, in the limit $\kappa,{m}>0$ and  $\pi \mathtt{k} R_0\gg 1$
\begin{equation}\label{m30}
\tan\left({(e^{\pi \mathtt{k} R_0}-1){m}\over 
\mathtt{k}}\right)={(e^{\pi \mathtt{k} R_0}-1)\mathtt{k}{m}\over 
(\mathtt{k}^2+{m}^2e^{\pi\mathtt{k} R_0})}
\end{equation}
This is the exact solution. Following  the 
same technique as in the 5D case, for $\pi \mathtt{k}R_0\gg 1$ we can 
neglect $Y_{3/2}$ for large arguments and reduce equation (\ref{m27}) 
to
\begin{equation}
J_{3\over 2}({{m}\over \mathtt{k}}e^{\pi \mathtt{k}R_0}) = 0\,.
\end{equation}
The asymptotic form of $\displaystyle J_\nu(x)=\sqrt{\frac{2}{nx^2}} \cos (x-\frac{\nu \pi}{2}-\frac {\pi}{4})$  is valid for any $\nu$ as long as $x\gg|\nu^2-1/4|$ 
is satisfied. Then, the basic idea is to set the argument of cosine 
function $(2N-1)\pi/2,\; N=1,2,3...,$ so that cosine vanishes. 
Thus, for arbitrary $\nu$ we have
\begin{equation}
  x-{\nu\pi\over 2}-{\pi\over 4} = (2N-1)\frac{\pi}{2}
\end{equation}
and get $x$ as
\begin{equation}
  x=\left(N+\frac{2\nu-1}{4}\right)\pi\,.
\end{equation}
This reduces to $x=(N+1/4)\pi$ for $\nu=1$ in 5D case. In 6D (or any higher dimensions), we obtain  
$\nu=3/2$ which instead gives $x=(N+1/2)\pi$ for the zeros of (2.14). (See the Numerical Analysis section for further details.) Putting  the argument 
of $J_{3/2}$ from above and using index $N$ for ${m}$, which is now discrete, we get
\begin{equation}\label{ybar}
  {m}_N=(N+\frac{1}{2})\pi \mathtt{k} e^{-\pi \mathtt{k} R_0}\,,\quad N=1,2,3...
\end{equation}
This is the form of ${m}$ which should enter 
the energy density equation.  Note that ${m}_N=0$ 
for $N=0$ , which will be important later. Also, for the 5D case  $\displaystyle {m}_N=(N+\frac{1}{4})\pi \mathtt{k} e^{-\pi \mathtt{k} R_0}\,,\quad N=1,2,3...$ \cite{Frank:2007jb}.

\section{The Casimir Force in the  RSI Model}

In RSI, there are two branes, localized at $y=0$ and $y=L$, with $Z_2$ symmetry $y\leftrightarrow -y$, $L+y\leftrightarrow L-y$. The 3-branes have equal opposite tensions. The positive tension brane has a fundamental scale $M_{\rm RS}$ and is hidden; SM fields are located on the negative tension brane, which is visible. The exponential warping factor gives rise to an effective scale on the visible brane located at $y=L$
\begin{equation}
M^2_{\rm Pl}=\frac{M_{\rm RS}^3} {\mathtt{k}}\left [1-e^{-2\mathtt{k}L}\right ]\,.
\end{equation}
A mechanism is needed to recover 4-dimensional General Relativity at low-energy, and this corresponds to introducing an extra degree of freedom known as the radion. A suitable choice of $L$ (often taken to be related to the compactified radius, $L=\pi R_0$) and $\mathtt{k}$ allows the Kaluza-Klein (KK) spectrum to be discrete, and the lowest masses to be of ${\cal O}$(TeV), which predicts different collider signatures at low energies from those of large extra dimensions. As consistency with the low energy theory requires  $M_{\rm Pl} \sim M_{\rm RS} \sim \mathtt{k}$, there are no additional hierarchies in this model. The scale of the 4-dimensional physical phenomena space transverse to the $5^{th}$ dimension is then specified by the warp factor : $\Lambda=M_{\rm Pl}\, e^{-\mathtt{k} \pi R_0}$, and if we take $\Lambda \sim 1$ TeV, then we expect $\mathtt{k}R_0 \sim12$.
\subsection {The 6D RSI Case}
The effective dispersion relation in 4D can be obtained from (\ref{eqn:4Deqn}) in $k$-space. It becomes
\begin{eqnarray}
\left((ik_\mu)(ik^\mu)+m_{eff}^2\right)X(k)=0\,,
\end{eqnarray} 
where $k_\mu=(w/c,\bf{k})$. Hence, the following dispersion relation is obtained
\begin{equation}
w(k)=c\sum_{n=0}^{\infty}\sqrt{{\bf k}^2+\frac{n^2}{R^2}+{m}_{eff}^2}\,,
\end{equation}
If we further assume two-plates configuration separated by a distance $a$ and apply Dirichlet boundary conditions along z-axis, we obtain the Casimir energy density per unit plate
\begin{equation}
{\mathbb E}_{6D}^I=V_{orb}\frac{\hbar}{2}\int \frac{d^2{\bf {k}_\perp}}{(2\pi)^2}\left(p\, c\sum_{n,l=0}^{\infty}\sqrt{{\bf k_\perp^2}+\frac{\pi^2 l^2}{a^2}+\frac{n^2}{R^2}+{m}^2}\right),
\end{equation}
where  $V_{orb}$ is the volume of the orbifold and $p$ is the polarization factor in 6D. This is our  starting point in calculating the Casimir force. 

\noindent After finding the discrete ${m}$ values corresponding to the RSI model with the above approximation, 
we can write the total frequency as
\begin{equation}
w = p \sum_{l,n,N=0}^{\infty^\prime}c \sqrt{{\bf k}_\bot^2+{\pi^2 
l^2\over a^2}+{n^2\over R^2}+{m}^2_N}
\label{eqn:3sum}
\end{equation}
where prime indicates that $l=n=N=0$ is excluded and 
${m}_N$ is given as
$$m_N= \left\{ \begin{array}{ll}
  0 &\mbox{ if $N=0$} \\
   \kappa(N+{1\over 2}) &\mbox{ if $N=1,2,\dots$}
        \end{array} \right.
$$
with $\kappa\equiv \pi \mathtt{k} e^{-\pi \mathtt{k} R_0}$. Note that the radius $R$ of the sixth dimension is in general 
different than $R_0$ of the fifth dimension.  Rewriting the triple 
sum  and rearranging the single and double sums to start with $0$, we obtain 
\begin{eqnarray}\label{m4.2}
w_{lnN}&=&pc\sum_{\ell,n,N=0}^{\infty^{\prime}} 
\sqrt{k^2_\perp+{\pi^2 \ell^2\over a^2}+{n^2\over 
R^2}+m_N^2}\nonumber\\
&=&pc\sum_{\ell,n=0}^{\infty^{\prime}} \sqrt{k^2_\perp+{\pi^2 
\ell^2\over a^2}+{n^2\over R^2}}+pc\sum_{\ell,n,N=0}^\infty 
\sqrt{k^2_\perp+{\pi^2 \ell^2\over a^2}+{n^2\over 
R^2}+m_N^2}\nonumber \\
&-&pc\sum_{\ell,n=0}^\infty \sqrt{k^2_\perp+{\pi^2 
\ell^2\over a^2}+{n^2\over R^2}+\frac{\kappa^2}{4}}. 
\end{eqnarray}
The Casimir energy density per unit plate is given by
\begin{equation}\label{m4.3}
{\mathbb E}_{6D}^I=V_{orbf}\frac{\hbar}{2}\int \frac{d^2{\bf 
{k}_\perp}}{(2\pi)^2}\left(p\, c\sum_{n,l=0}^{\infty}\sqrt{{\bf 
k_\perp^2}+\frac{\pi^2 l^2}{a^2}+\frac{n^2}{R^2}+{m}_N^2}\right).
\end{equation}
 We obtain
\begin{eqnarray*}
{\mathbb E}_{6D}^I
&=&V_{orbf}{pc\hbar\Gamma(-{3\over 2})\over 8\pi \Gamma(-{1\over 
2})}\bigg(\sum_{\ell,n=0}^{\infty^{\prime}} \left({{\pi^2 
\ell^2\over a^2}+{n^2\over R^2}}\right)^{3\over 
2}+\sum_{\ell,n,N=0}^\infty \left({{\pi^2 \ell^2\over a^2}+{n^2\over 
R^2}+\kappa^2(N+{1\over 2})^2}\right)^{3\over 
2}\nonumber \\
&-&\sum_{\ell,n=0}^\infty \left({{\pi^2 \ell^2\over a^2}+{n^2\over 
R^2}+{\kappa^2\over 4}}\right)^{3\over 2}\bigg). 
\end{eqnarray*}
In order to evaluate this expression we introduce the Epstein Zeta Function $Z_N(a_1,\dots,a_N;s)$ \cite{Elizalde:1994gf}:
\begin{equation} 
Z_N(a_1 \dots,a_N;s)=\sum_{n_1,\dots,n_N=-\infty}^{\infty^\prime}\left( a_1n_1^2+ \dots a_N n_N^2\right)^{-\frac{s}{2}}
\end{equation}
and the inhomogeneous Epstein Hurwitz Function $E_N^c$ \cite{Elizalde:1994gf}:
\begin{equation} 
E_N^c(s;a_1, \dots, a_N;c_1, \dots, c_N)=\sum_{n_1,\dots,n_N=0}^{\infty}\left [ \sum_{j=1}^{N} a_j\left( n_j+c_j\right)^{2}+c \right]^{-s}.
\end{equation}

\noindent We can write the energy density between the plates in a compact form as
\begin{eqnarray}\label{m4.6}
{\mathbb E}_{6D}^I&=&V_{orb}{pc\hbar\Gamma(-{3\over 2})\over 8\pi 
\Gamma(-{1\over 2})}\!\Bigg\{\!\! \sum_{\ell,n=0}^{\infty^{\prime} }
\left({{\pi^2 \ell^2\over a^2}+{n^2\over R^2}}\right)^{3\over 
2}+E_3(-{3\over 2};{\pi^2\over a^2},{1\over R^2},\kappa^2;0,0,{1\over 
2})-E_2^{\kappa^2\over 4}(-{3\over 2};{\pi^2\over a^2},{1\over 
R^2};0,0)\Bigg\}\nonumber\\
&=&V_{orb}{pc\hbar\Gamma(-{3\over 2})\over 8\pi 
\Gamma(-{1\over 2})}
\bigg({1\over 4}Z_2({1\over R},{\pi\over a};-3)+{1\over 4}Z_1({\pi 
\over a};-3)+{1\over 4}Z_1({1\over R};-3)+
E_3(-{3\over 2};{\pi^2\over a^2},{1\over R^2},\kappa^2;0,0,{1\over 
2})\nonumber\\
&-&E_2^{\kappa^2\over 4}(-{3\over 2};{\pi^2\over a^2},{1\over R^2};0,0)\bigg)
\end{eqnarray}
where the last form is obtained computing the explicit analytic expansion for the sums.
Using the reflection formula for the Epstein Zeta function \cite{Ambjorn:1981xv}
\begin{equation}
\label{eq:reflection}
Z_p(a_1,\dots,a_p;s)=a_1^{-1}\dots a_p^{-1}{\pi^s\over 
\pi^{p/2}}{\Gamma({p-s\over 2})\over \Gamma({s\over 2})}Z_p({1\over 
a_1},\dots,{1\over a_p};p-s),
\end{equation}
 we may write the first double sum on the right side
\begin{eqnarray*}
&&{1\over 4}\bigg[Z_2({1\over R},{\pi\over a};-3)+
Z_1({1\over R};-3)+Z_1({\pi \over a};-3)\bigg]=\frac{1}{2 \Gamma(-\frac32) \pi^{7/2}}\bigg[ \frac {3\,Ra}{8 \pi}Z_2\left( R,\frac{a}{\pi}; 5\right) +\frac{1}{R^3} \zeta(4)+\frac{\pi^3}{a^3} \zeta(4) \bigg]
\end{eqnarray*}
where in the last term we introduced the Riemann Zeta Function \cite{Elizalde:1994gf}: $\zeta(s)=\sum_{n=1}^{\infty}n^{-s}$. 

\noindent The triple sum $ E_3(-{3\over 2};{\pi^2\over a^2},{1\over 
R^2},\kappa^2;0,0,{1\over 2})$ is evaluated using \cite{Elizalde:1994gf}
\begin{eqnarray}\label{m5.7}
E_3(s;a_1,a_2,a_3;c_1,c_2,c_3)
&=&{1\over \Gamma(s)}\sum_{m=0}^\infty {(-1)^m\over 
m!}a_1^m\zeta_H(-2m,c_1)\Gamma(s+m)E_2(s+m;a_2,a_3;c_2,c_3)\nonumber\\
&+&{1\over 2}\sqrt{\pi\over a_1}{\Gamma(s-{1\over 2})\over 
\Gamma(s)}E_2(s-{1\over 2};a_2,a_3;c_2,c_3)\nonumber\\
&+&\sqrt{\pi\over a_1}{\cos(2\pi c_1)\over 
\Gamma(s)}\sum_{n_1=1}^\infty\sum_{n_2,n_3=0}^\infty \int_0^\infty 
dt~ t^{s-{3\over 2}}\exp[-{\pi^2 n_1^2\over a_1 
t}-t\sum_{j=2}^3a_j(n_j+c_j)^2]\nonumber
\end{eqnarray}
to obtain
\begin{eqnarray}\label{m5.9}
E_3(-{3\over 2};{\pi^2\over a^2},{1\over R^2},&\kappa^2&;0,0,{1\over 
2})=\zeta_H(0,0)E_2(-{3\over 2};{1\over R^2},\kappa^2;0,{1\over 
2})+{1\over 2}{a\over \sqrt{\pi}}{\Gamma(-2)\over \Gamma(-{3\over 
2})}E_2(-2;{1\over R^2},\kappa^2;0,{1\over 2})\nonumber\\
&+&{2\over a\sqrt{\pi}\Gamma(-{3\over 
2})}\sum_{\ell=1}^\infty\sum_{n,N=0}^\infty
\ell^{-2}
\left({n^2\over R^2}+\kappa^2(N+{1\over 2})^2\right) K_{2}\left({2a 
\ell}\sqrt{{n^2\over R^2}+\kappa^2(N+{1\over 2})^2}\right).\nonumber
\end{eqnarray}

\noindent The double sum
$\displaystyle E_2^{\kappa^2/4} \left(-\frac32; \frac{\pi^2}{a^2},\frac{1}{R^2};0,0\right)  $
can be regarded as a special case of the inhomogeneous 
Epstein-Hurwitz Zeta function and its expansion \cite{Elizalde:1994gf}
\begin{eqnarray}\label{m5.13}
E_2^c(s;a_1,a_2;c_1,c_2)
&=& {a_2^{-s}\over \Gamma(s)}\sum_{m=0}^\infty 
{(-1)^m\Gamma(s+m)\over m!}\left({a_1\over 
a_2}\right)^m\zeta_H(-2m,c_1)E_1^{c/a_2}(s+m;1;c_2)\nonumber\\
&+&{a_2^{1/2-s}\over 2}\sqrt{\pi\over a_1}{\Gamma(s-{1\over 2})\over 
\Gamma(s)}E_1^{c/a_2}(s-1/2;1;c_2)\nonumber\\
&+&{2\pi^s\over \Gamma(s)}\cos(2\pi 
c_1)a_1^{-s/2-1/4}\sum_{n_1=1}^\infty \sum_{n_2=0}^\infty 
n_1^{s-1/2}[a_2(n_2+c_2)^2+c]^{-s/2+1/4}\nonumber\\
&\times&K_{s-1/2}\left({2\pi n_1\over 
\sqrt{a_1}}\sqrt{a_2(n_2+c_2)^2+c}\right),
\end{eqnarray}
where $K_\nu (x)$ is the modified Bessel function $\displaystyle K_\nu(x)=\frac{\pi}{2}\frac{(I_{-\nu}-I_{\nu})}{\sin \nu \pi}$, with $\displaystyle I_\nu(x)=\sum_{s=0}^\infty \frac{1}{s!(s+\nu)!}\left( \frac{x}{2}\right)^{2s+\nu}$. The sum  
is finally obtained as 
\begin{eqnarray}\label{m5.18}
E_2^{\kappa^2/4} \left(-\frac32; \frac{\pi^2}{a^2},\frac{1}{R^2};0,0\right)  &=&
\zeta_H(0,0)E_1^{\kappa^2\over 4}(-{3\over 2};{1\over R^2},0)+{1\over 
2}{a\over \sqrt{\pi}}{\Gamma(-2)\over \Gamma(-{3\over 
2})}E_1^{\kappa^2\over 4}(-2;{1\over R^2};0)\nonumber\\
&+&{2\over a\sqrt{\pi}\Gamma(-{3\over 
2})}\sum_{\ell=1}^\infty\sum_{n=0}^\infty \ell^{-2}\left({n^2\over 
R^2}+{\kappa^2\over 4}\right)K_{2}\left(2\ell a\sqrt{{n^2\over 
R^2}+{\kappa^2\over 4}}\right).
\end{eqnarray}

\noindent Combining the three terms given above, we obtain, for the Casimir energy between the plates: \begin{eqnarray}\label{m5.20}
\mathbb E_{6D}^I&=&V_{orb}{pc\hbar\Gamma(-{3\over 2})\over 8\pi 
\Gamma(-{1\over 2})}\Bigg\{
{1\over 2\Gamma(-{3\over 2}){\pi^{7\over 2}}}\bigg[{3\over 
8}{Ra\over\pi}Z_2(R,{a\over \pi};5)+
{1\over R^3}\zeta(4)+{\pi^3\over a^3}\zeta(4)\bigg]\nonumber\\
&+&\zeta_H(0,0)E_2(-{3\over 2};{1\over R^2},\kappa^2;0,{1\over 
2})+{1\over 2}{a\over \sqrt{\pi}}{\Gamma(-2)\over \Gamma(-{3\over 
2})}E_2(-2;{1\over R^2},\kappa^2;0,{1\over 2})\nonumber\\
&+&{2\over a\sqrt{\pi}\Gamma(-{3\over 
2})}\sum_{\ell=1}^\infty\sum_{n,N=0}^\infty
\ell^{-2}
\left({n^2\over R^2}+\kappa^2(N+{1\over 2})^2\right) K_{2}\left({2a 
\ell}\sqrt{{n^2\over R^2}+\kappa^2(N+{1\over 2})^2}\right)\nonumber\\
&-&\zeta_H(0,0)E_1^{\kappa^2\over 4}(-{3\over 2};{1\over 
R^2},0)-{1\over 2}{a\over \sqrt{\pi}}{\Gamma(-2)\over \Gamma(-{3\over 
2})}E_1^{\kappa^2\over 4}(-2;{1\over R^2};0)\nonumber\\
&-&{2\over a\sqrt{\pi}\Gamma(-{3\over 
2})}\sum_{\ell=1}^\infty\sum_{n=0}^\infty \ell^{-2}\left({n^2\over 
R^2}+{\kappa^2\over 4}\right)K_{2}\left(2\ell a\sqrt{{n^2\over 
R^2}+{\kappa^2\over 4}}\right)\Bigg\}.
\end{eqnarray}

\noindent Before proceeding further the energy 
without plates must be subtracted. Let's call the energy density without the plates $\epsilon_{\rm np}^{6D}$. The renormalized 
total energy shift can be given as
\begin{equation}\label{m6.1}
  E_{\rm 6D}^{\rm I\,(ren)} = \mathbb E_{6D}^I A-\epsilon_{\rm np}^{6D} A\ a,
\end{equation}
where $A$ is the area of one plate. The energy density without  
plates $\epsilon_{\rm np}^{6D}$ is given by
\begin{eqnarray}\label{m6.2}
  \epsilon_{\rm np}^{6D}&=&V_{orb}{\hbar\over
2}\int{d^3\textbf{k}\over (2\pi)^3}\left(pc\sum_{n,N=0\atop n=N\neq 
0}^{\infty^\prime}
\sqrt{\textbf{k}^2+{n^2\over R^2}+m_N^2}\right)\nonumber \\
&=&V_{orb}{\hbar\over
2}pc\int{d^3\textbf{k}\over (2\pi)^3}\left(\sum_{n=1}^{\infty}
\sqrt{\textbf{k}^2+{n^2\over R^2}}+\sum_{n=0}^\infty
\sum_{N=0}^\infty \sqrt{\textbf{k}^2+{n^2\over 
R^2}+m_N^2}-\sum_{n=0}^\infty\sqrt{\textbf{k}^2+{n^2\over 
R^2}+{\kappa^2\over 4}}\right).
\end{eqnarray}
This yields, in terms of the previously defined functions
\begin{eqnarray}\label{m6.6}
&&\epsilon^{6D}_{\rm np}
=-\frac {V_{orb}\hbar pc}{
16} \Bigg[ {{3\over 4\pi^6}}{1\over R^4}
\zeta(5)+\frac{\Gamma(-2)}
{\pi^{3\over 2}\Gamma(-{1\over 2})} E_2(-2;{1\over R^2},\kappa^2;0,{1\over 2})
-\frac{\Gamma(-2)}
{\pi^{3\over 2}\Gamma(-{1\over 2})} E_1^{\kappa^2\over 4}(-2;{1\over R^2},0)\Bigg].\nonumber
\end{eqnarray}
Then the total energy in the 6D RSI is 
\begin{eqnarray}\label{m7.5}
E_{6D}^{\rm I\,(ren)}&=&A V_{orb}{pc\hbar\Gamma(-{3\over 2})\over 8\pi 
\Gamma(-{1\over 2})}\Bigg\{
{1\over 2\Gamma(-{3\over 2}){\pi^{7\over 2}}}\bigg[{3\over 
8}{Ra\over\pi}Z_2(R,{a\over \pi};5)+
{1\over R^3}\zeta(4)+{\pi^3\over a^3}\zeta(4)\bigg]\nonumber\\
&+&{2\over a\sqrt{\pi}\Gamma(-{3\over 
2})}\sum_{\ell=1}^\infty\sum_{n,N=0}^\infty
\ell^{-2}
\left({n^2\over R^2}+\kappa^2(N+{1\over 2})^2\right) K_{2}\left({2a 
\ell}\sqrt{{n^2\over R^2}+\kappa^2(N+{1\over 2})^2}\right)\nonumber\\
&-&{2\over a\sqrt{\pi}\Gamma(-{3\over 
2})}\sum_{\ell=1}^\infty\sum_{n=0}^\infty \ell^{-2}\left({n^2\over 
R^2}+{\kappa^2\over 4}\right)K_{2}\left(2\ell a\sqrt{{n^2\over 
R^2}+{\kappa^2\over 4}}\right) \Bigg\}\nonumber\\
&+&{Aa V_{orb}\over 2}{\hbar pc\over
16{\pi}^{5}}{{3\over 4\pi}}{1\over R^4}
\zeta(5) +({\rm terms~ independent~of~a}).
\end{eqnarray}
We note that this expression contains no divergent terms and is finite, as expected, because the divergent terms in $\epsilon_{\rm np}^{6D}$ cancel exactly the ones in $\mathbb E_{6D}^I$.
The Casimir force is given by the derivative of the Casimir 
energy $E_{6D}^{\rm I\,(ren)}$ with respect to the plate separation $a$:
\begin{eqnarray}\label{m8.1}
F_{6D}^{I}&=&-{\partial  E_{6D}^{\rm I\,(ren)}\over \partial a}.
\end{eqnarray}
We use the the definition of $Z_2$ to calculate derivatives.  The force becomes, 
 using (\ref{m7.5}),
\begin{eqnarray}\label{m8.6}
{\mathbb F_{6D}^{I}\over A}
&=&V_{orb}{pc\hbar\over 16\pi^2}\Bigg\{{1\over 
2{\pi^{3}}}\Bigg[{3\over 8}{R\over \pi} Z_2(R,{a\over \pi};5)
-{15\over 8}{Ra^2\over \pi^3} 
\sum_{n,N=-\infty}^{\infty'}N^2\left({n^2 R^2}+{a^2N^2\over 
\pi^2}\right)^{-{7\over 2}}\!\!\!\!-{3\pi^3\over 
a^{4}}\zeta(4)\nonumber\\
&-&{{3\over 4\pi}}{1\over R^4} \zeta(5)\Bigg]-{6\over 
a^2}\sum_{\ell,N=1}^\infty\sum_{n=0}^\infty
\ell^{-2}
\left({n^2\over R^2}+\kappa^2(N+{1\over 2})^2\right) K_{2}\left({2a 
\ell}\sqrt{{n^2\over R^2}+\kappa^2(N+{1\over 2})^2}\right)\nonumber\\
&-&{4\over a}\sum_{\ell,N=1}^\infty\sum_{n=0}^\infty
\ell^{-1}
\left({n^2\over R^2}+\kappa^2(N+{1\over 2})^2\right)^{3\over 
2}K_{1}\left({2a \ell}\sqrt{{n^2\over R^2}+\kappa^2(N+{1\over 
2})^2}\right)\Bigg\}
\end{eqnarray}
where we use the identity $\displaystyle \partial_zK_{\nu}(z)=-K_{\nu-1}(z)-\frac{\nu}{z}K_\nu(z)$. This expression is our final formula for the Casimir force density (force per unit area) for the 6 dimensional RSI model. Comparison with the  expression for the force in 5D models \cite{Frank:2007jb} shows that the terms in the square bracket arise as contributions from the extra compact dimension and are called UED-like terms, as in \cite{Poppenhaeger:2003es}; while the last two terms given as  Bessel functions are the contribution of the 5D warped coordinates and will be called Bessel terms. 
In the limit $R \rightarrow \infty$: 
 \begin{eqnarray}\label{m8.16}
{\mathbb F_{5D}^{I}\over A}
&=&-V_{orb}{pc\hbar\over 16\pi^2}  \Bigg\{ {3\over 
2a^{4}}\zeta(4) +\Bigg[\frac{6 \kappa^2}{a^2}
\sum_{N=0}^\infty\sum_{\ell=1}^\infty
\ell^{-2}
(N+{1\over 2})^2K_{2}\left({2a 
\ell}\kappa(N+{1\over 2})\right)\nonumber\\
&+&4 a \kappa \sum_{N=0}^\infty\sum_{\ell=1}^\infty
\ell^{-1}
(N+{1\over 2})^3K_{1}\left({2a \ell}\kappa(N+{1\over 
2})\right) \Bigg]  \Bigg\}.
\end{eqnarray}
The first term is the Casimir force without the extra dimensions,  and the two Bessel terms coincide with those in the 5D consideration \cite{Frank:2007jb}, except for the argument of the Bessel function. The difference in arguments arises from the location of the zeros of the Bessel functions in 5 and 6 dimensions, as shown in the previous section, and $N+1/2$ has to be replaced with $N+1/4$ as well, since the solution for $m_N$ in 5D is different from the one in 6D. Note that $\displaystyle -V_{orb} \frac{3 pc \hbar }{15 \pi^2 a^4} \zeta(4)=-\frac{p \pi^2c \hbar}{480 a^4}=\frac {p}{2} \mathbb F_{\rm no RS}$. 
Thus the expressions (2.18) from  \cite{Frank:2007jb} and  (\ref{m8.16})   are consistent with each other. 

\subsection{The $q$D RSI case}

We continue by generalizing this expression to $q$ extra compact dimensions. We proceed as before. We calculate first the energy density between the plates, then the energy density without the plates. We use this to evaluate the energy shift and then compute the force.

The energy density between the plates for $q$ compact dimensions ($q$D) is 
\begin{eqnarray}\label{m11.1}
w_{qD}&=&pc\sum_{\ell=0}^\infty \prod_{i=1}^q\sum_{n_i,N=0
}^{\infty^{\prime} }\sqrt{\textbf{k}_\bot^2+{\pi^2 
\ell^2\over a^2}+\sum_{j=1}^q{n_j^2\over R_j^2}+m_N^2}\nonumber\\
&=&pc\sum_{\ell=0}^\infty \prod_{i=1}^q\sum_{n_i=0 
}^{\infty^{\prime}} \sqrt{\textbf{k}_\bot^2+{\pi^2 
\ell^2\over a^2}+\sum_{j=1}^q{n_j^2\over R_j^2}}+pc\sum_{\ell=0}^\infty 
\prod_{i=1}^q\sum_{n_i,N=0}^\infty\sqrt{\textbf{k}_\bot^2+{\pi^2 
\ell^2\over a^2}+\sum_{j=1}^q{n_j^2\over R_j^2}+m_N^2}\nonumber\\
&-&pc\sum_{\ell=0}^\infty \prod_{i=1}^q\sum_{n_i=0}^\infty 
\sqrt{\textbf{k}_\bot^2+{\pi^2 \ell^2\over 
a^2}+\sum_{j=1}^q{n_j^2\over R_j^2}+{\kappa^2 \over 4}}
\end{eqnarray}
which give the energy density as
\begin{eqnarray}\label{m11.2}
{\mathbb E}_{qD}^I&=&V_{orb}{\hbar\over 
2}\int{d^{2}\textbf{k}_\bot\over 
(2\pi)^{2}}\left(pc\sum_{\ell=0}^\infty 
\prod_{i=1}^q \sum_{n_i,N=0}^{\infty^\prime} 
\sqrt{\textbf{k}_\bot^2+{\pi^2 \ell^2\over 
a^2}+\sum_{j=1}^q {n_j^2\over R_j^2}+m_N^2}\right)\nonumber\\
&=&V_{orb}{pc \hbar\over  8\pi}{\Gamma(-{3\over 
2})\over \Gamma(-{1\over 2})}\Bigg(
\sum_{\ell=0}^\infty \prod_{i=1}^q\sum_{n_i=0}^{\infty^\prime}({\pi^2 \ell^2\over 
a^2}+\sum_{j=1}^q{n_j^2\over R_j^2})^{{3\over 2}}+\sum_{\ell=0}^\infty \prod_{i=1}^q\sum_{n_i,N=0}^\infty({\pi^2 
\ell^2\over a^2}+\sum_{j=1}^q{n_j^2\over R_j^2}+\kappa^2(N+{1\over 
2})^2)^{{3\over 2}}\nonumber\\
&-&\sum_{\ell=0}^\infty \prod_{i=1}^q\sum_{n_i=0}^\infty({\pi^2 
\ell^2\over a^2}+\sum_{j=1}^q{n_j^2\over R_j^2}+{\kappa^2\over 
4})^{{3\over 2}}\bigg).
\end{eqnarray}
We obtain, after expressing everything in terms of Epstein Zeta and Epstein Hurwitz functions:
\begin{eqnarray}\label{m11.13}
{\mathbb E}_{qD}^I&=&V_{orb}{pc \hbar\over  8\pi}{\Gamma(-{3\over 
2})\over \Gamma(-{1\over 2})}
\Bigg(
{1\over 2^{q+1}}\bigg[Z_{q+1}({\pi\over a},{1\over R_1},\dots,{1\over 
R_q};-3)+\sum_qZ_q(\{{\pi\over a},{1\over R_1},\dots,{1\over 
R_q}\};-3)\nonumber\\
&+&\sum_{q-1}Z_{q-1}(\{{\pi\over a},{1\over R_1},\dots,{1\over 
R_q}\};-3)+\sum_{q-2}Z_{q-2}(\{{\pi\over a},{1\over 
R_1},\dots,{1\over R_q}\};-3)\nonumber\\
&+&\dots+\sum_{1}Z_{1}(\{{\pi\over a},{1\over R_1},\dots,{1\over R_q}\};-3)
\bigg]-{1\over 2}E_{q+1}(-{3\over 2};\underbrace{{1\over 
R_1^2},\dots,{1\over R_q^2},\kappa^2}_{\mbox{$q+1$ 
terms}};\underbrace{0,0,\dots,{1\over 2}}_{\mbox{$q+1$ 
terms}})\nonumber\\
&+&{1\over 2}{a\over \sqrt{\pi}}{\Gamma(-2)\over \Gamma(-{3\over 2})}
E_{q+1}(-2;\underbrace{{1\over R_1^2},\dots,{1\over 
R_q^2},\kappa^2}_{\mbox{$q+1$ terms}};\underbrace{0,0,\dots,{1\over 
2}}_{\mbox{$q+1$ terms}})\nonumber\\
&+&{2\over a\sqrt{\pi}\Gamma(-{3\over 
2})}\sum_{\ell=1}^\infty\sum_{n_1,n_2,\dots,n_{q},N=0}^\infty 
\ell^{-2}\left(\sum\limits_{i=1}^q {n_i^2\over 
R_i^2}+\kappa^2(N+{1\over 2})^2\right)K_{2}\left(2 
a\ell\sqrt{\sum\limits_{i=1}^q {n_i^2\over R_i^2}+\kappa^2(N+{1\over 
2})^2}\right)
\nonumber\\
&+&{1\over 2}E_{q}^{\kappa^2\over 4}(-{3\over 2};\underbrace{{1\over 
R_1^2},\dots,{1\over R_{q}^2}}_{\mbox{$q$ 
terms}};\underbrace{0,0,\dots,0}_{\mbox{$q$ terms}})-{1\over 
2}{a\over \sqrt{\pi}}{\Gamma(-2)\over \Gamma(-{3\over 
2})}E_{q}^{\kappa^2\over 4}(-2;\underbrace{{1\over 
R_1^2},\dots,{1\over R_{q}^2}}_{\mbox{$q$ 
terms}};\underbrace{0,0,\dots,0}_{\mbox{$q$ terms}})\nonumber\\
&-&{2\over a\sqrt{\pi}\Gamma(-{3\over 
2})}\sum_{\ell=1}^\infty\sum_{n_1,n_2,\dots,n_q}^\infty
\ell^{-2}\left(\sum\limits_{i=1}^q {n_i^2\over R_i^2}+{\kappa^2\over 
4}\right)K_{2}\left(2a\ell\sqrt{\sum\limits_{i=1}^q {n_i^2\over 
R_i^2}+{\kappa^2\over 4}}\right) \Bigg)
\end{eqnarray}
where $\sum_q$ indicates the sum over a permutation of the set $q$-terms of 
$\{{\pi\over a},{1\over R_1},\dots,{1\over R_q}\}$, $\sum_{q-1}$ 
indicates the sum over a permutation of $(q-1)$-terms of $\{{\pi\over 
a},{1\over R_1},\dots,{1\over R_q}\}$, and so on. 
Note that $\sum_{1}Z_{1}$ is the sum over a permutation of one term 
of $\{{\pi\over a},{1\over R_1},\dots,{1\over R_q}\}$.

The energy density without the plate $\epsilon_{\rm np}^{qD}$ is given by
\begin{eqnarray}\label{m12.1}
\epsilon_{\rm np}^{qD}&=&V_{orb}{\hbar\over  2}\int{d^{3}\textbf{k}\over (2\pi)^{3}}pc 
\prod_{i=1}^q\sum_{n_i=0}^{\infty^{\prime}} 
\sqrt{\textbf{k}^2+\sum_{j=1}^q{n_j^2\over R_j^2}}+V_{orb}{\hbar\over  2}\int{d^{3}\textbf{k}\over (2\pi)^{3}}
pc \prod_{i=1}^q\sum_{n_i=0,N=0}^\infty 
\sqrt{\textbf{k}^2+\sum_{j=1}^q{n_j^2\over R_j^2}+m_N^2}\nonumber\\
&-&V_{orb}{\hbar\over  2}\int{d^{3}\textbf{k}\over (2\pi)^{3}}
pc \prod_{i=1}^q\sum_{n_i=0}^\infty 
\sqrt{\textbf{k}^2+\sum_{j=1}^q{n_j^2\over R_j^2}+{\kappa^2 \over 4}}
\end{eqnarray}
which gives:
\begin{eqnarray}\label{m12.4}
\epsilon_{\rm np}^{qD}&=&V_{orb}{pc\hbar\Gamma(-2)\over 
2^{q+3}\pi^{3\over 2}\Gamma(-{1\over 2})}
\bigg[Z_{q}(\{{1\over R_1},\dots,{1\over 
R_q}\};-4)+\sum_{q-1}Z_{q-1}(\{{1\over R_1},\dots,{1\over 
R_q}\};-4)\nonumber\\
&+&\sum_{q-2}Z_{q-2}(\{{1\over R_1},\dots,{1\over R_q}\};-4)+
\dots+\sum_{1}Z_{1}(\{{1\over R_1},\dots,{1\over R_q}\};-4)\bigg]\nonumber\\
&+&V_{orb}{pc\hbar\Gamma(-2)\over  16\pi^{3\over 2}\Gamma(-{1\over 
2})}E_{q+1}(-2;{1\over R_1^2},\dots,{1\over 
R_q^2},\kappa^2;0,\dots,{1\over 2})\nonumber\\
&-&V_{orb}{pc\hbar\Gamma(-2)\over  16\pi^{3\over 2}\Gamma(-{1\over 
2})}E_{q}^{\kappa^2\over 4}(-2;{1\over R_1^2},\dots,{1\over 
R_q^2};0,\dots,0).
\end{eqnarray}
 The renormalized energy shift is then given by
\begin{equation}\label{m13.1}
E_{qD}^{\rm I\,(ren)}=\mathbb E_{qD}^IA-\epsilon_{\rm np}^{qD}Aa.
\end{equation}
As before, the divergent terms in the vacuum energy without plates exactly cancel the divergent terms in the energy density between the plates. This yields, for the energy density, in a compact form
\begin{eqnarray}\label{m13.6}
\frac{{ E}_{qD}^{\rm I\,(ren)}}{A}&=&\frac{V_{orb}\,pc \hbar} 
{\Gamma(-{1\over 2})}
\Bigg\{
\frac {1}{ 2^{q+3}\pi^{4}}
\sum\limits_{k=1}^{q+1}{\Gamma({k+3\over 
2})\over\pi^{{k\over2}}}\bigg[\sum_{\langle k \rangle}\{{a \over 
\pi}, R_1,...,R_q\}_{\langle k \rangle}Z_k(\{{a\over 
\pi},R_1,...,R_q\}_{\langle k \rangle};k+3)\nonumber\\
&-&{a\over \pi}\sum_{\langle k-1 \rangle}\{{a \over \pi}, 
R_1,...,R_q\}_{\langle k-1 \rangle}Z_{k-1}(\{{a\over 
\pi},R_1,...,R_q\}_{\langle k-1 \rangle};k+3)
\bigg]\nonumber\\
&+&{1\over {4a\pi^{3\over 2}}}\sum_{\ell=1}^\infty  \ell^{-2} \bigg[  \!\!\!\!
\sum_{n_1,n_2,\dots,n_{q}=0}^\infty\sum_{N=1}^\infty 
\left(\sum\limits_{i=1}^q {n_i^2\over R_i^2}+\kappa^2(N+{1\over 
2})^2\right)K_{2}\left(2 a\ell\sqrt{\sum\limits_{i=1}^q {n_i^2\over 
R_i^2}+\kappa^2(N+{1\over 2})^2}\right)\bigg] \Bigg\}
\nonumber\\
&+&({\rm terms~ independent ~of~ a})
\end{eqnarray}
\noindent where $\{{a \over \pi}, R_1,...,R_q\}_{\langle k \rangle}$ 
represents all possible sets with the $k$ element chosen in each case. 
So, $\{{a \over \pi}, R_1,...,R_q\}_{\langle 0 \rangle}$ is an empty 
set.
In the formula above we highlighted the finiteness of the energy density by using the reflection 
formula, see \cite{Ambjorn:1981xv} and (\ref{eq:reflection}), 
 allowing us to write:
\begin{eqnarray*}
Z_{q}({1\over R_1},\dots,{1\over R_q};-4)&=&{1\over \pi^{{q\over 
2}+4}}{\Gamma({q+4\over 2})\over 
\Gamma(-2)}\left(\prod_{i=1}^qR_i\right)Z_q(R_1,\dots,R_q;q+4)\\
\sum_{q-1}Z_{q-1}(\{{1\over R_1},\dots,{1\over R_q}\};-4)&=&{1\over 
\pi^{{q-1\over 2}+4}} {\Gamma({q+3\over 2})\over 
\Gamma({-2})}\sum_{q-1}\{{R_1},\dots,{R_q}\}Z_{q-1}(\{{ R_1},\dots,{ 
R_q}\};q+3).
\end{eqnarray*}
For calculating the force, $\displaystyle \frac{{\mathbb 
F}_{qD}^{I}}{A}=-\frac{ \partial (\frac{{ 
E}_{qD}^{\rm I\,(ren)}}{A})}{\partial a}$, note that using
$$Z_{q+1}({a\over 
\pi},R_1,\dots,R_q;q+4)=\sum_{\ell,n_1,n_2,\dots,n_q=-\infty}^{\infty'}\left({a^2\ell^2\over 
\pi^2}+R_1^2 n_1^2+\dots+R_q^2 n_q^2\right)^{-{q+4\over 2}}$$
we obtain
\begin{eqnarray*}
{\partial \over \partial a}Z_{q+1}({a\over 
\pi},R_1,\dots,R_q;q+4)
&=&{-{(q+4)a\over 
\pi^2}}\sum_{\ell,n_1,n_2,\dots,n_q=-\infty}^{\infty'}\ell^2\left({a^2\ell^2\over 
\pi^2}+R_1^2 n_1^2+\dots+R_q^2 n_q^2\right)^{-{q+6\over 2}}.
\end{eqnarray*}
Further, for
$\displaystyle 
{\partial \over \partial a}\sum_q {\{{a\over \pi},R_1,\dots,R_q\}\over
\pi^{{q\over 2}+3}}Z_q(\{{a\over \pi},R_1,\dots,R_q\};q+3)$, 
we note that only one term that will be independent of $a$. Thus, we may write it as
\begin{eqnarray*}
&&{\partial \over \partial a}\sum_{q} {\{{a\over \pi},R_1,\dots,R_q\}\over
\pi^{{q\over 2}}}Z_q(\{{a\over \pi},R_1,\dots,R_q\};q+3)=\sum_{q-1} 
{\{R_1,\dots,R_q\}\over
\pi^{{q\over 2}+1}}Z_q(\{{a\over \pi},R_1,\dots,R_q\};q+3)\\
&-&\frac{(q+3)a^2}{ \pi^2}\sum_{q-1} {\{R_1,\dots,R_q\}\over
\pi^{{q\over 
2}+1}}\sum_{l,\{n_1,\dots,n_q\}=-\infty}^{\infty'}l^2\left({a^2l^2\over 
\pi^2}+\{R_1^2n_1^2+\dots+R_q^2n^2_q\}\right)^{-{q+5\over 2}}.
\end{eqnarray*}
Then we obtain the general expression for the force in $q$D RSI, in compact form
\begin{eqnarray}\label{m13.8}
\frac{{\mathbb F}^{I}_{qD}}{A}
&=&\frac{V_{orb}\,pc \hbar}  {2\sqrt{\pi}}
\Bigg\{
\frac {1}{ 2^{q+3}\pi^{5}}\Bigg[
\sum\limits_{k=1}^{q+1}{\Gamma({k+3\over 2})\over\pi^{{k\over2}}}
\sum_{\langle k-1 \rangle}\{R_1,...,R_q\}_{\langle k-1 
\rangle}\Bigg(Z_k({a\over \pi},\{R_1,...,R_q\}_{\langle k-1 
\rangle};k+3)\nonumber\\
&-& Z_{k-1}(\{R_1,...,R_q\}_{\langle k-1 
\rangle};k+3)-{(k+3)a^2\over\pi^2}\sum\limits_{\ell,\{n_i\}_{\langle 
k-1 
\rangle}=-\infty}^{\infty^\prime}\ell^2\bigg({a^2\ell^2\over\pi^2}+\{R_i 
n_i\}^2_{\langle k-1 \rangle}\bigg)^{-{k+5\over 2}}\Bigg)\Bigg]
\nonumber\\
&-&{1\over \pi^{3\over 2}}\sum_{\ell=1}^\infty 
\sum_{n_1,n_2,\dots,n_{q}=0}^\infty\sum_{N=1}^\infty 
\bigg[\frac{3}{4a^2\ell^2} \left(\sum\limits_{i=1}^q {n_i^2\over 
R_i^2}+\kappa^2(N+{1\over 2})^2\right)K_{2}\left(2 
a\ell\sqrt{\sum\limits_{i=1}^q {n_i^2\over R_i^2}+\kappa^2(N+{1\over 
2})^2}\right)
\nonumber\\
&+&\frac{1}{2a\ell}
\left(\sum\limits_{i=1}^q  {n_i^2\over R_i^2}+\kappa^2(N+{1\over 
2})^2\right)^{\frac32}K_{1}\left(2 a\ell\sqrt{\sum\limits_{i=1}^q 
{n_i^2\over R_i^2}+\kappa^2(N+{1\over 2})^2}\right)
\bigg] \Bigg\}.
\end{eqnarray}
In differentiation we always assume that, other 
than ${a\over \pi}$ term, there is always at least one term chosen 
from the rest,  $\{R_1, R_2,...,R_q\}$. 
 We include the term with $k=1$ in the sum with the convention that $\{R_1, R_2,...,R_q\}_{\langle 0\rangle}=1$ and $Z_0=0$.

As a check on our result, we verify for the case $q=1$ it reproduces the 6D force found in the previous section. We have, for $q=1$
\begin{eqnarray}\label{m13.9}
\frac{{\mathbb F}^{I}_{6D}}{A}
&=&\frac{V_{orb}\,pc \hbar}  {2\sqrt{\pi}}
\Bigg\{
\frac {1}{ 2^{4}\pi^{5}}\Bigg[
{\Gamma({5\over 2})\over\pi}
R \Bigg(Z_2({a\over \pi},R; 5)-
 Z_1(R ; 5)-{5a^2\over\pi^2}\sum\limits_{\ell,n=-\infty}^{\infty^\prime}\ell^2\bigg({a^2\ell^2\over\pi^2}+n^2R^2\bigg)^{-{5\over 2}}\Bigg)\nonumber\\
&+&  \frac{1}
{\pi^{1\over 2}}\bigg(
Z_{1}({a\over \pi};4)
-\frac{4a^2}{\pi^2}\!\!\!\sum_{\ell=-\infty}^{\infty'}\ell^2 
\left({a^2\ell^2\over \pi^2}\right)^{-3} \bigg)
\Bigg]
\nonumber\\
&-&{1\over \pi^{3\over 2}}\sum_{\ell, N=1}^\infty 
\sum_{n=0}^\infty
\bigg[\frac{3}{4a^2\ell^2} \left( {n^2\over 
R^2}+\kappa^2(N+{1\over 2})^2\right)K_{2}\left(2 
a\ell\sqrt{ {n^2\over R^2}+\kappa^2(N+{1\over 
2})^2}\right)
\nonumber\\
&+&\frac{1}{2a\ell}
\left(  {n^2\over R^2}+\kappa^2(N+{1\over 
2})^2\right)^{\frac32}K_{1}\left(2 a\ell\sqrt{
{n^2\over R^2}+\kappa^2(N+{1\over 2})^2}\right)
\bigg] \Bigg\}.
\end{eqnarray}
Observing that $\displaystyle 
Z_{1}({a\over \pi};4)=-\frac{6 \pi^3}{a^4} \zeta(4)$, 
the expression (\ref{m13.9}) is identical to (\ref{m8.6}). Thus the $q$D result coincides with the 6D in the appropriate limit.

\section{The Casimir Force in RSII models}
\label{sec4}

In the Randall Sundrum model II (RSII), there is only one positive tension brane. It may be thought of as a limiting case of the previous model, where one brane is located at infinity ($R_0 \rightarrow \infty$). If there is a warped extra dimension, large scales at the Planck brane are shifted at the TeV brane and the relationship between energy scales is given by
\begin{equation}
M^2_{\rm Pl}=\frac{M_{\rm RS}^3} {\mathtt{k}}\,.
\end{equation}
Limits on Newton's law set lower bounds on the brane tension and the fundamental scale of RSII \cite{Maartens:2003tw}. It has been shown also \cite{Giddings:2000mu} that massless bulk scalars in the Randall
 Sundrum background (2.1) have similar properties as gravitons: there exists a
localized zero mode and a Kaluza-Klein continuum of arbitrarily light states
weakly interacting with matter residing on the brane.
The spectrum of RSII is a continuous spectrum of $m> 0$ KK modes, and there are no ${\cal O}$(TeV) signatures for this model at the colliders. The infinite dimension makes a finite contribution to the 5-dimensional volume because of the warp factor, and the effective size of the extra dimension probed is $1/\mathtt{k}$. 
\subsection{The 6D RSII case}
We use the same procedure to evaluate the force in the RSII model  as in the 5D case. The difference here is that the spectrum is continuous in ${m}$ and consists of all $m \ge 0$. The extra dimensional summation in $N$ becomes an integral with measure $dm/\mathtt{k}$. We note that in RSII in addition 
to the continuous mode, there is the zero (discrete) mode which needs 
to be handled separately. The total energy density is the sum of zero 
and continuous modes:
\begin{equation}
 \mathbb E^{II}_{6D,\rm tot}=\mathbb E^{II}_{6D(0)} +\mathbb {E}^{II}_{6D}.
\end{equation}
where
\begin{eqnarray}\label{eq9.1}
\mathbb E^{II}_{6D(0)} &=& {cp\hbar\over 2}\int{d^2\textbf{k}_\bot\over 
(2\pi)^2}\left(\sum_{l,n=0}^{\infty^{\prime}}
\sqrt{\textbf{k}_\bot^2+{\pi^2 l^2\over a^2}+{n^2\over
R^2}}\right)\nonumber\\
\mathbb E^{II}_{6D}&=&
\frac {c p\hbar }
{2} \int \frac{dm}{\mathtt{k}}\int{d^2\textbf{k}_\bot\over 
(2\pi)^2}\left(\sum_{l,n=0}^{\infty^{\prime}}
\sqrt{\textbf{k}_\bot^2+{\pi^2 l^2\over a^2}+{n^2\over
R^2}+{m}^2}\right)
\end{eqnarray}
where the first expression is the discrete zero-mode and the second is the continuous $m>0$ mode.  We deal with the zero-mode first. $\mathbb E^{II}_{6D(0)}$ is already calculated in RSI. It can be 
written as    
\begin{eqnarray}\label{m5.3}
\mathbb E^{II}_{6D(0)}&=&{pc\hbar\Gamma(-{3\over 2})\over 8\pi \Gamma(-{1\over 2})}
\bigg({1\over 4}Z_2({1\over R},{\pi\over a};-3)+{1\over 4}Z_1({\pi 
\over a};-3)+{1\over 4}Z_1({1\over R};-3)\bigg).
\end{eqnarray}
We re-write the zero mode energy density by using the 
reflection formula (\ref{eq:reflection}) to give
\begin{eqnarray}\label{m5.3a}
\mathbb E^{II}_{6D(0)}&=&-{pc\hbar \over 64\pi^5}
\bigg({3\over 4}{R a \over \pi }Z_2(R,{a\over \pi};5)+{2\over 
R^3}(1+{\pi^3R^3\over a^3})\zeta(4)\bigg).
\end{eqnarray}
We calculate the no-plate form of the zero-mode energy as 
well. The expression is the same as the zero-mass term in the no-plate energy density in the 6D RSI model:
\begin{equation}
\epsilon_{\rm np}^{6D(0)}=-{1\over 4}{ pc\hbar\over
16{\pi}^{5}}{{3\over 4\pi}}{2\over R^4}
\zeta(5).
\end{equation}
The energy shift for the zero mode becomes (including the 
area $A$ and volume ($aA$) factors for the energy)
\begin{equation}
  { E^{{\rm II\,(ren)}}_{6D(0)} \over A}=-{pc\hbar \over 64\pi^5}
\bigg({3\over 4}{R a \over \pi }Z_2(R,{a\over \pi};5)+{2\over 
R^3}(1+{\pi^3R^3\over a^3})\zeta(4)-{3\over 4\pi}{2a\over 
R^4}\zeta(5)\bigg)
\end{equation}
yielding the force for the zero mode component
\begin{eqnarray}
{\mathbb {F}^{II}_{6D(0)}\over A} =
{pc\hbar \over 64\pi^5}
\bigg\{{3\over 4}{R \over \pi }Z_2(R,{a\over \pi};5)-{15\over 4}{Ra^2 
\over \pi^3 }\sum\limits_{n_1,n_2=-\infty}^{\infty^{\prime}} 
n_1^2\left({a^2n_1^2\over 
\pi^2}+{R^2n_2^2}\right)^{-7/2}-{6\pi^3\over a^4}\zeta(4)-{3\over 
4\pi}{2\over R^4}\zeta(5)\bigg\}.
\end{eqnarray}
The continuos 
mode energy density (4.3) can further be written as
\begin{eqnarray*}
\mathbb {E}^{II}_{6D}
=
\frac {c p\hbar }
{2} \int \frac{dm}{\mathtt{k}}\int{d^2\textbf{k}_\bot\over 
(2\pi)^2}\bigg(\sum_{l,n=1}^{\infty}
\sqrt{\textbf{k}_\bot^2+{\pi^2 l^2\over a^2}+{n^2\over
R^2}+{m}^2}+\sum_{n=1}^{\infty}
\sqrt{\textbf{k}_\bot^2+{n^2\over
R^2}+{m}^2}+\sum_{l=1}^{\infty}
\sqrt{\textbf{k}_\bot^2+{\pi^2 l^2\over a^2}+{m}^2}\bigg)
\end{eqnarray*}
yielding, for the energy between the plates,
\begin{eqnarray}
\label{m9.7}
\mathbb {E}^{II}_{6D}&=&\frac {c p\hbar }
{64} \frac {\Gamma(-2)} {\mathtt{k}
\sqrt{\pi} \Gamma(-{1\over 2}) } \left[{2aR\over 
\pi^6\Gamma(-2)}Z_2({a\over \pi},R;6)+\left({1\over R^4}+{\pi^4\over 
a^4}\right) {3\over 2\pi^{4}\Gamma(-2)}\zeta(5)\right] \nonumber\\
&=&\frac {c p\hbar }
{32}\frac{1}{\mathtt{k}
\sqrt{\pi}\Gamma(-{1\over 2})}\left[{aR\over \pi^6}Z_2({a\over 
\pi},R;6)+{3\over 4\pi^{4}}\left({1\over R^4}+{\pi^4\over 
a^4}\right)\zeta(5)\right].
\end{eqnarray}
The energy without the plate in RSII models is simply, for continuous modes,
\begin{eqnarray}\label{m9.8}
{\epsilon}^{6D}_{\rm np}&=&
\frac {c p\hbar }
{2} \int \frac{dm}{\mathtt{k}}\int{d^3\textbf{k} \over 
(2\pi)^3}\left(\sum_{n=1}^{\infty}
\sqrt{\textbf{k} ^2+{n^2\over
R^2}+{m}^2}\right)\nonumber\\
&=&-\frac {c p\hbar }
{64\pi R^5 \mathtt{k}{(2\sqrt{\pi})}} {2\over \pi^{11\over 2}}2\zeta(6)-\frac {c p\hbar }
{32\pi^7 R^5 \mathtt{k}} \zeta(6).
\end{eqnarray}
We now compute the force in Casimir 6D including the 
zero mode contribution as
\begin{eqnarray}\label{m10.1}
{\mathbb {F}^{II}_{6D}\over A}&=&{\mathbb {F}^{II}_{6D(0)}\over A} 
-{\partial {E}^{\rm II\,(ren)}_{6D}\over \partial a}= {\mathbb {F}^{II}_{6D(0)}\over A} -\frac {\partial (\mathbb {E}^{II}_{6D}- A a 
{\epsilon}^{6D}_{\rm np})}{\partial a} \nonumber\\
&=&{\mathbb {F}^{II}_{6D(0)}\over A} +\frac {c p\hbar }
{64}{1\over {\mathtt{k}}
{\pi}}\left[{R\over \pi^6}Z_2({a\over \pi},R;6)-{6a^2R\over 
\pi^8}\!\!\!\sum\limits_{n_1,n_2=-\infty}^{\infty^{\prime}} \!\!\!n_1^2\left({a^2n_1^2\over 
\pi^2}+{R^2n_2^2}\right)^{-4}\!\!\!
-{3\over a^5}\zeta(5)-\frac {2 }
{\pi^6 R^5} \zeta(6) \right].
\end{eqnarray}
\noindent This is our final expression for the 6D force in RSII models, where
\begin{eqnarray}
\displaystyle
&& {{F}^{II}_{6D(0)}\over A} =
{pc\hbar \over 64\pi^5}
\bigg\{{3\over 4}{R \over \pi }Z_2({a\over \pi},R;5)-{15\over 4}{Ra^2 
\over \pi^3 }\!\!\!\sum\limits_{n_1,n_2=-\infty}^{\infty} \!\!\!
n_1^2\left({a^2n_1^2\over 
\pi^2}+{R^2n_2^2}\right)^{-7/2}\!\!\!-{6\pi^3\over a^4}\zeta(4)-{3\over 
4\pi}{2\over R^4}\zeta(5)\bigg\}.\nonumber
\end{eqnarray}
Comparing this expression with the corresponding one in 5D, we have, for $R \rightarrow \infty$
\begin{eqnarray}\label{m10.3}
{\mathbb {F}^{II}_{5D}\over A}&=&-{cp\hbar \over 64\pi^5}\bigg[ {6\pi^3\over a^4}\zeta(4)\bigg] -\frac {c p\hbar }
{64}{1\over {\mathtt{k}}
{\pi}}\left[
{3\over a^5}\zeta(5) \right]
\nonumber\\
&=&\mathbb F_{\rm no RS}\left (\frac{p}{2}+\frac{45p}{2\pi^3a \mathtt{k}}\zeta(5) \right)
\end{eqnarray}
where in the last expression we used $\displaystyle \mathbb F_{\rm noRS}=-\frac{6c\hbar}{32 \pi^2 a^4}\zeta(4)=-\frac{\pi^2c \hbar}{240 a^4} $
which is exactly what we obtained in (2.20) of  \cite{Frank:2007jb}.

\subsection{The $q$D RSII case}

 As we did in 6D, we quote the zero mode renormalized 
energy and the corresponding force from $q$D RSI results. We 
can get them from (\ref{m13.6}) and (\ref{m13.8}), 
respectively. It is sufficient to switch off the term with Hurwitz Zeta 
functions. We have
\begin{eqnarray}\label{m14.1a}
\frac{{ E}^{\rm II\,(ren)}_{qD(0)}}{A}&=&\frac{pc \hbar} 
{\Gamma(-{1\over 2})}
\frac {1}{ 2^{q+4}\pi^{4}}
\sum\limits_{k=1}^{q+1}{\Gamma({k+3\over 
2})\over\pi^{{k\over2}}}\bigg[\sum_{\langle k \rangle}\{{a \over 
\pi}, R_1,...,R_q\}_{\langle k \rangle}Z_k(\{{a\over 
\pi},R_1,...,R_q\}_{\langle k \rangle}; k+3)\nonumber\\
&-&{a\over \pi}\sum_{\langle k-1 \rangle}\{{a \over \pi}, 
R_1,...,R_q\}_{\langle k-1 \rangle}Z_{k-1}(\{{a\over 
\pi},R_1,...,R_q\}_{\langle k-1 \rangle};k+3)\bigg]
+({\rm terms~ independent ~of~ a}).
\end{eqnarray}
and
  \begin{eqnarray}\label{m14.1b}
\frac{{\mathbb F}^{II}_{qD(0)}}{A}
&=&\frac{pc \hbar}  {2\sqrt{\pi}}
\frac {1}{ 2^{q+4}\pi^{5}}
\sum\limits_{k=1}^{q+1}{\Gamma({k+3\over 2})\over\pi^{{k \over 2}}}
\sum_{\langle k-1 \rangle}\{R_1,...,R_q\}_{\langle k-1 
\rangle}\Bigg(Z_k({a\over \pi},\{R_1,...,R_q\}_{\langle k-1 
\rangle};k+3)\nonumber\\
&-& Z_{k-1}(\{R_1,...,R_q\}_{\langle k-1 
\rangle};k+3)-{(k+3)a^2\over\pi^2}\sum\limits_{\ell,\{n_i\}_{\langle 
k-1 \rangle}=-\infty}^{\infty^\prime}\ell^2\bigg({a^2\ell^2\over\pi^2}+\{R_i 
n_i\}^2_{\langle k-1 \rangle}\bigg)^{-{k+5\over 2}}\Bigg).
\end{eqnarray}

\noindent Calculation of the energy and the force for 
the continuous mode gives
\begin{eqnarray}\label{eq14.1}
{\mathbb E}^{II}_{qD}&=&
\frac {c p\hbar }
{2} \int \frac{dm}{\mathtt{k}}\int{d^2\textbf{k}_\bot\over 
(2\pi)^2}\left(\sum_{\ell=0}^\infty \prod_{i=1}^q\sum_{n_i=0 
}^{\infty^\prime}
\sqrt{\textbf{k}_\bot^2+{\pi^2 l^2\over a^2}+\sum_{j=1}^q{n_j^2\over
R_j^2}+{m}^2}\right)\nonumber\\
&= &
{pc\hbar\over{2^{q+5}\mathtt{k}\sqrt{\pi}\pi^4}}{1\over\Gamma({-{1\over 2}})}
\sum\limits_{k=1}^{q+1}{\Gamma({k+4\over 2})\over\pi^{{k\over2}}}
\sum_{\langle k \rangle}\{{a\over \pi},R_1,...,R_q\}_{\langle k 
\rangle}Z_k(\{{a\over \pi},R_1,...,R_q\}_{\langle k \rangle};k+4).
 \end{eqnarray}
 For the energy without the plate in $q$D for RSII models, we have
\begin{eqnarray}\label{eq14.9}
{\epsilon}_{\rm np}^{qD}&=&
\frac {c p\hbar }
{2} \int \frac{dm}{\mathtt{k}}\int{d^3\textbf{k} \over (2\pi)^3}\left(\prod_{i=1}^q\sum_{n_i=0}^{\infty^\prime}
\sqrt{\textbf{k} ^2+\sum_{j=1}^q \frac{n_j^2}{
R_j^2}+{m}^2}\right)
\end{eqnarray}
which yields
  \begin{eqnarray}\label{eq14.13}
{\epsilon}^{qD}_{\rm np}&=&\frac {c p\hbar }
{2^{q+4}\pi \mathtt{k}\Gamma(-{1\over 2})}
\bigg[{\left(\prod\limits_{i=1}^q{R_i}\right)\over \pi^{{q\over 
2}+5}}{\Gamma({q+5\over 2})}Z_q({R_1,\dots,R_q};q+5)\nonumber\\
&+&\sum_{q-1}
{\{{R_1},\dots,{R_q}\}\over \pi^{{q-1\over 2}+5}}{\Gamma({q+4\over 
2})}Z_{q-1}(\{R_1,\dots,R_q\};q+4)\\
&+&\sum_{q-2}
{\{{R_1},\dots,{R_q}\}\over \pi^{{q-2\over 2}+5}}{\Gamma({q+3\over 
2})}Z_{q-2}(\{R_1,\dots,R_q\};q+3)+\dots+2\sum_{1}
{\{{R_1},\dots,{R_q}\}\over \pi^{{11\over 2}}}Z_{1}(\{R_1,\dots,R_q\};6)
\bigg].\nonumber
\end{eqnarray}
Following the same method as used for the force in $q$D RSI model, we 
obtain the force
$$\frac{{\mathbb F}^{II}_{qD}}{A}=\frac{{\mathbb 
F}^{II}_{qD(0)}}{A}-\frac{\partial E_{qD}^{\rm II\,(ren)} }{\partial 
a}=\frac{{\mathbb F}^{II}_{qD(0)}}{A}-\frac{\partial ({E}^ 
{II}_{qD} -Aa{\epsilon}^{qD}_{np})}{\partial a}.$$
Then the general expression for the force in $q$D RSII models is, in compact form
 \begin{eqnarray}\label{m14.15}
\frac{{\mathbb F}_{qD}^{II}}{A}
&=&\frac{{\mathbb F}^{II}_{qD(0)}}{A}+\frac{pc \hbar}  { 2^{q+6}\pi^6 \mathtt{k}}
\Bigg\{
\sum\limits_{k=1}^{q+1}{\Gamma({k+4\over 2})\over\pi^{{k\over2}}}
\sum_{\langle k-1 \rangle}\{R_1,...,R_q\}_{\langle k-1 
\rangle}\Bigg(Z_k({a\over \pi},\{R_1,...,R_q\}_{\langle k-1 
\rangle};k+4)\nonumber\\
&-&Z_{k-1}(\{R_1,...,R_q\}_{\langle k-1 
\rangle};k+4)-{(k+4)a^2\over\pi^2}\sum\limits_{\ell,\{n_i\}_{\langle 
k-1 
\rangle}=-\infty}^{\infty^\prime}\ell^2\bigg({a^2\ell^2\over\pi^2}+\{R_i 
n_i\}^2_{\langle k-1 \rangle}\bigg)^{-{k+6 \over 2}}\Bigg)
\Bigg\}
\end{eqnarray}
 where $\mathtt{k}$ in the factor is the curvature of the space and 
the force corresponding to zero mode is (by reproducing the expression from 
(\ref{m14.1b}) )
  \begin{eqnarray}
\frac{{\mathbb F}^{II}_{qD(0)}}{A}
&=&\frac{pc \hbar}  {2\sqrt{\pi}}
\frac {1}{ 2^{q+4}\pi^{5}}
\sum\limits_{k=1}^{q+1}{\Gamma({k+3\over 2})\over\pi^{{k\over2}}}
\sum_{\langle k-1 \rangle}\{R_1,...,R_q\}_{\langle k-1 
\rangle}\Bigg[Z_k({a\over \pi},\{R_1,...,R_q\}_{\langle k-1 
\rangle};k+3)\nonumber\\
&-&Z_{k-1}(\{R_1,...,R_q\}_{\langle k-1 
\rangle};k+3)-{(k+3)a^2\over\pi^2}\sum\limits_{\ell,\{n_i\}_{\langle 
k-1 
\rangle}=-\infty}^{\infty^\prime}\ell^2\bigg({a^2\ell^2\over\pi^2}+\{R_i 
n_i\}^2_{\langle k-1 \rangle}\bigg)^{-{k+5\over 2}}\Bigg].
\end{eqnarray}
As a check on our result, we verify that in the case $q=1$ it reproduces the RSII 6D force found in the previous section. We have, for $q=1$
\begin{eqnarray}\label{m14.2}
\frac{{\mathbb F}^{II}_{6D}}{A}
&=&
\frac {pc \hbar}{ 2^{7}\pi^{6}\mathtt{k} }\Bigg\{
{\Gamma({3})\over\pi^{1}}
R \Bigg(Z_2({a\over \pi},R; 6)-
 Z_1(R ; 6)-{6a^2\over\pi^2}\sum\limits_{\ell,n=-\infty}^{\infty^\prime}\ell^2\bigg({a^2\ell^2\over\pi^2}+n^2R^2\bigg)^{-{4}}\Bigg]\nonumber\\
&+&  \frac{\Gamma(5/2)}
{\pi^{1\over 2}}\bigg [
Z_{1}({a\over \pi};5)
-\frac{5a^2}{\pi^2}\!\!\!\sum_{\ell=-\infty}^{\infty'}\ell^2 
\left({a^2\ell^2\over \pi^2}\right)^{-{7\over 2}} \bigg] \Bigg\}\nonumber \\
&+&\frac{pc \hbar}  {2\sqrt{\pi}}
\frac {1}{ 2^{4}\pi^{5}}\Bigg\{
{\Gamma({5\over 2})\over\pi^{1}} R
\Bigg[ Z_2({a\over \pi},R; 5)-
 Z_1(R ; 5)-{5a^2\over\pi^2}\sum\limits_{\ell,n=-\infty}^{\infty^\prime}\ell^2\bigg({a^2\ell^2\over\pi^2}+n^2R^2\bigg)^{-{7\over 2}}\Bigg ]\nonumber\\
&+&  \frac{1}
{\pi^{1\over 2}}\bigg[
Z_{1}({a\over \pi};4)
-\frac{4a^2}{\pi^2}\!\!\!\sum_{\ell=-\infty}^{\infty'}\ell^2 
\left({a^2\ell^2\over \pi^2}\right)^{-3} \bigg]
\Bigg \}.
\end{eqnarray}
Observing again that $\displaystyle 
Z_{1}({a\over \pi};4)=-\frac{6 \pi^3}{a^4} \zeta(4)$, $\displaystyle -\frac{\Gamma(5/2)}
{\pi^{1\over 2}}\bigg(Z_{1}({a\over \pi};5)
+\frac{5a^2}{\pi^2}\!\!\!\sum_{\ell=-\infty}^{\infty'}\ell^2 
\left({a^2\ell^2\over \pi^2}\right)^{-{7\over 2}} \bigg)= -\frac34 \frac{8 \pi^5}{a^5} \zeta(5) $, as well as $\displaystyle  R Z_1(R;6)=\frac{2}{R^5}\zeta(6)$, 
the expression (\ref{m14.2}) is identical to (\ref{m10.1}). We have thus shown that the $q$D expression for the Casimir force reduces, with the appropriate substitutions, to the previously obtained expression in 6D.

\section{Numerical Analysis and Discussion}
In the previous two sections we presented complete compact expressions for the Casimir force between two parallel planes in Randall Sundrum Models extended by $q$ compact dimensions. The models thus contain one non-compact extra dimension $y$, associated with a curvature parameter $\kappa$ which is determined by the 5D mass, as well as $q$ compact ones. In the simplest case, the model will contain one compact ($\theta$)   and one non-compact dimension (the 6D case). We analyze the 6D model first, then comment on an arbitrary  number of compact dimensions by comparing 7D with 6D.
\begin{figure}[b]
\begin{center}
\hspace*{-2cm}
	\includegraphics[width=5in]{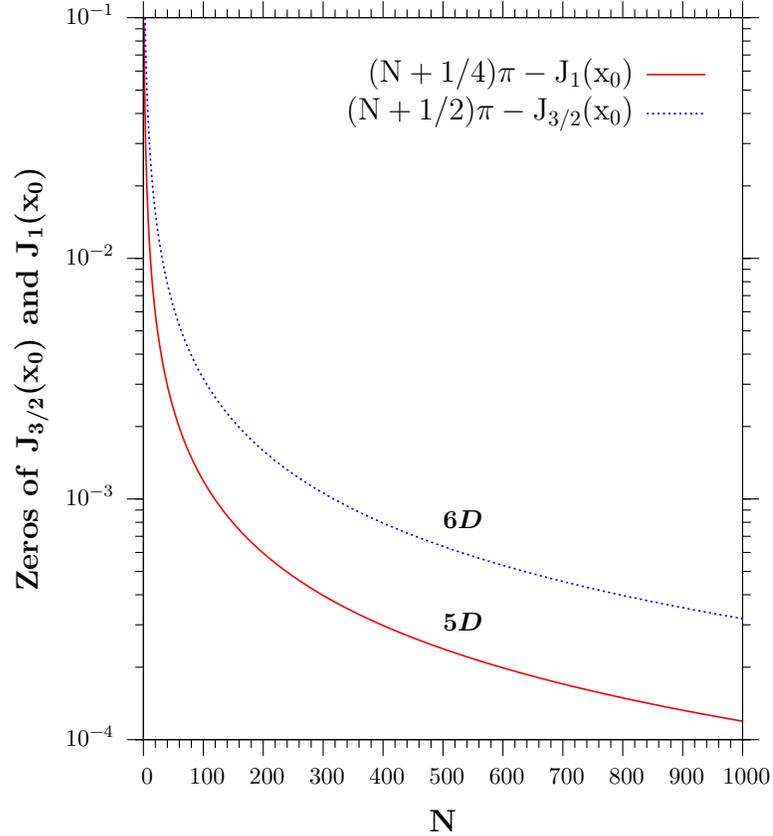} 
\end{center}
\vskip -0.2in
\caption{The approximations used for the zeros of the Bessel functions of first kind $J_1(x)$ and $J_{3/2}(x)$ for 5D and 6D, respectively. }\label{approx}
\end{figure}

Additionally we distinguish between the two Randall Sundrum scenarios.  In the (phenomenologically) more popular version of the model there are two branes, localized at $y=0$ and $y=\pi R_0$, with opposite tensions, called RSI. In the numerical analysis we concentrate on the RSI model, as RSII does not set any limits on the number or size of extra dimensions. The only approximation we have used in the derivation of the force is the form of $m_N$. We obtained  $m_N$ as $(N+1/2)\kappa$ in 6D by taking $(N+1/2)\pi$ as the zeros of Bessel function $J_{3/2}(x_0)$. In 5D the relevant Bessel function   was $J_1(x_0)$ and $m_N$ was approximated as $(N+1/4)\kappa$ in our previous study \cite{Frank:2007jb}. In Fig.~\ref{approx} we show the difference between the exact zeros of $J_{1,3/2}$ and our approximation as a functions of $N$. As seen, the approximation works better as $N$ increases, but is already good for small $N$ values as well.  

The contribution of the compact dimensions ($\theta_1, \theta_2,...$) to the force resemble the one in universal extra dimensions (UED), not surprisingly, as they enter into the dispersion relation in the same way. Whereas a full expression for the Casimir force for an arbitrary number of universal extra dimensions does not exist in the literature, our 6D result is the same as in  \cite{Poppenhaeger:2003es} in the limit $\kappa \rightarrow 0$. We note that our force expression for $q$D naturally recovers the $q$-dimensional generalization of 5D UED as $\kappa\to 0$.

The difference between the Casimir force in RSI and RSII models rests in the appearance of the Bessel function terms, which are the contribution from the discrete KK spectrum, absent in RSII due to the lack of the second boundary condition.  The Casimir force was calculated for RSII model in both 6D and $q$D case in \cite{Linares:2007yz} and in \cite{Linares:2008am}, respectively, using the Green's function method. We find it rather difficult to compare our results to theirs, as they do not present an expression for the force in terms of analytic functions. In the 6D analysis, they obtain the same effective 4D Casimir force as we do  ($\displaystyle \frac{p}{2} F_{\rm noRS}$) under appropriate limits. Comparing their results in $q$D to ours, we both have $\displaystyle \mathbb {F}^{II}_{qD}=F_{\rm noRS}\left ( 1+ \Delta_{RS} \right)$, but while in our case $\Delta_{RS}$ is a dimensionless term, as it should be, their term is of the form $\left[\rm (const)/Length\right]$, which is clearly problematic.

\begin{figure}[htb]
\begin{center}$
\hspace*{-1cm}
	\begin{array}{cc}
	\includegraphics[width=3.8in]{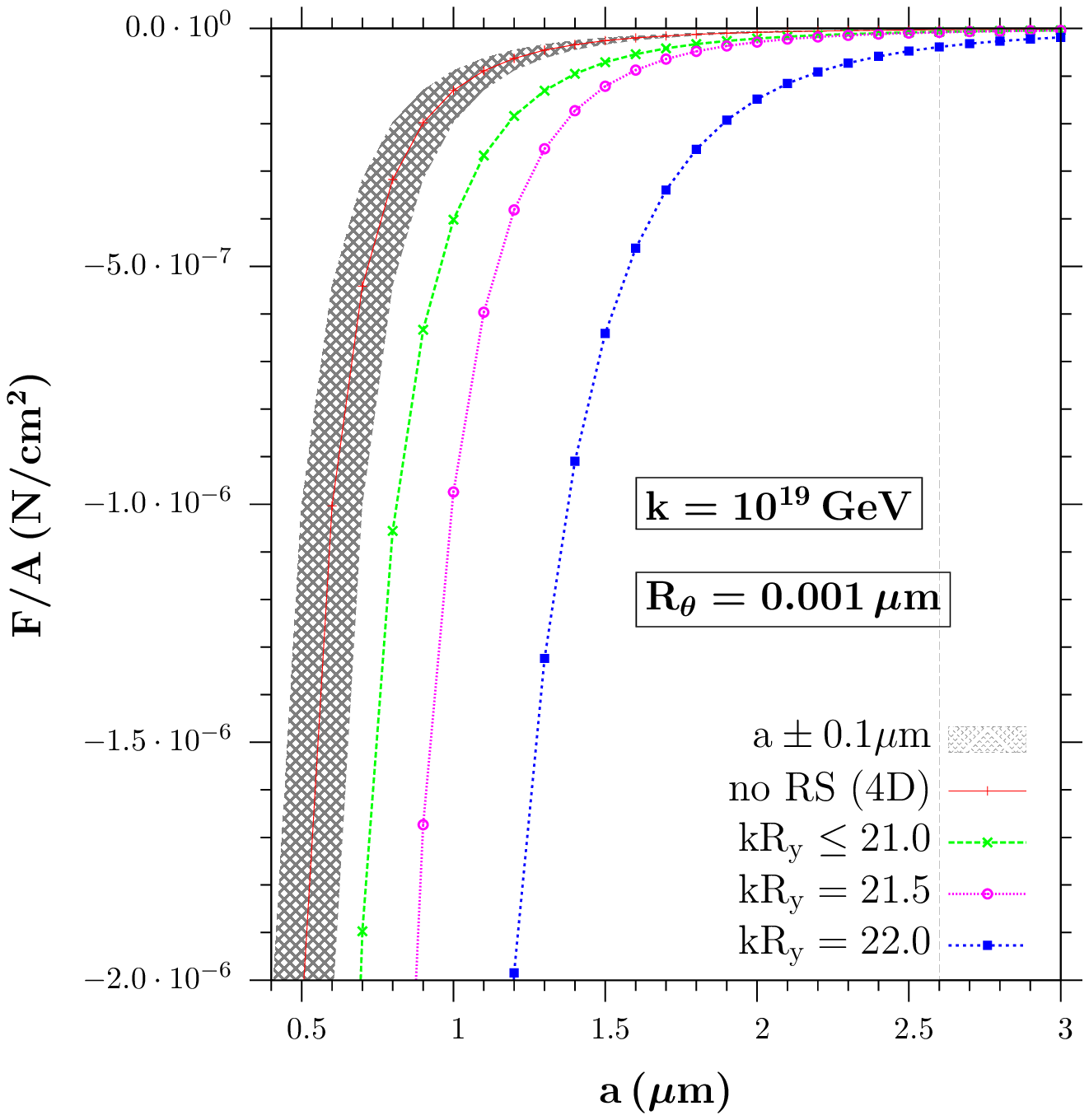} &
\hspace*{-0.9cm}
	\includegraphics[width=3.8in]{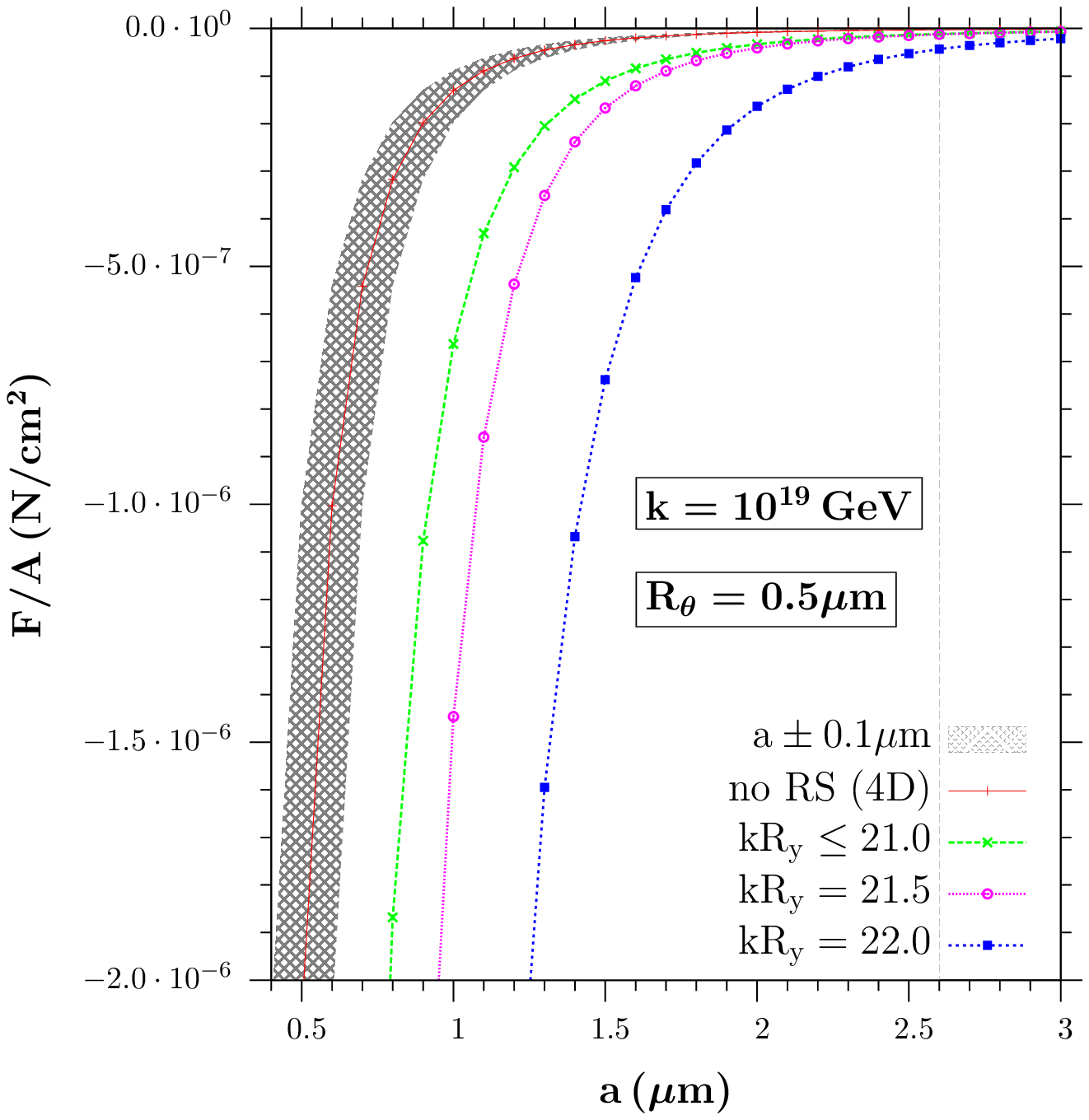}
	\end{array}$	
\end{center}
\vskip -0.2in
\caption{The Casimir force in $6D$ as a function of plate separation $a$ for various $\mathtt{k}R_y$ values with $\mathtt{k}=10^{19}$ GeV and $R_\theta=0.001 \,\mu m$ in the left panel and $R_\theta=0.5 \,\mu m$ in the right panel. The shaded region represents $\pm 0.1\,\mu m$ error for the measurement of plate separation $a$.}\label{6d}
\end{figure}

We now proceed to analyze the implications of the expressions obtained on the size and number of the extra dimensions in RSI model.  We first analyze the effect of one extra compact dimension in Fig.~\ref{6d} as a function of plate separation $a$. We fix $\mathtt{k}=10^{19}$ GeV and the radius of $\theta$ as $R_\theta=0.001\,\mu$m in the left panel and $R_\theta=0.5\,\mu$m in the right panel. In each case, there are curves for various $\mathtt{k}R_y$ values. As we can see, all curves with $\mathtt{k}R_y$ values less than 21  coincide and are represented as one curve. In this case there is no contribution coming from RS (Bessel terms) and our results reduce to those in to 5D UED. This feature can also be seen from our analytical expression given in (\ref{m8.6}).
The RS effects are included only in the Bessel function terms and the rest is the same as for 5D UED for the 6D case and for $(q-1)$D for the $q$D RS case. We analyzed  (\ref{m8.6}) more systematically  using an analytical approach. This method is also helpful when estimating what one would expect numerically  by adding more extra dimensions. 

We first tried to understand why  effects appear from RS at around $\mathtt{k}R_y=21$. We concentrate on the term with the modified Bessel function $K_2$ (a similar analysis will hold for the $K_1$ term as well):\begin{equation}
 \sum_{\ell,N=1}^\infty\sum_{n=0}^\infty
\ell^{-2}
\left({n^2\over R^2}+\kappa^2(N+{1\over 2})^2\right) K_{2}\left({2a 
\ell}\sqrt{{n^2\over R^2}+\kappa^2(N+{1\over 2})^2}\right).
\end{equation}
The dominant contributions come when the argument of $K_2$ is small, thus the first few terms of series are dominant. In addition, as $R_\theta$ gets smaller and smaller, it increases the argument of $K_2$  (for non-zero $n$ of course,) and no matter how small $\kappa$ is, there is no contribution from terms with Bessel functions. Thus, the $n=0$ term in the series is dominant. For simplicity,  take $\ell=N=1$ and call the argument of $K_2$ as $x_{n,\ell,N}$. Then, the dominant term is $x_{0,1,1}=a\kappa$. The limiting form of $K_\alpha(x)$  is, for $0\ll x\ll\sqrt{\alpha+1}$
\begin{equation}
 K_\alpha(x)\longrightarrow \frac{\Gamma(\alpha)}{2}\left(\frac{2}{x}\right)^\alpha.
\end{equation}
The condition on $x_{0,1,1}$ gives a critical value where the Bessel terms start being sizable. For example, for $a=0.5\,\mu$m and $\mathtt{k}=10^{19}$ GeV, the critical value of $\mathtt{k}R_y$ becomes $(\mathtt{k}R_y)_C=21.0034$. At any value greater than this critical value, $\kappa$ becomes  exponentially smaller and the Bessel  terms completely dominate contributions from the other terms. This is exactly the effect seen in Fig.~\ref{6d}. The same analysis for the $K_1$ term gives 21.068 for the critical value of $\mathtt{k}R_y$. The limiting form of the $K_1$ and $K_2$ terms, including the factors in front (but not the sign) becomes $(\kappa/a)^2$ and $3/a^4$, respectively.

We need to emphasize another feature. Like the Casimir force in 4D, the force in 6D vanishes as the plate separation increases.  The $\zeta(5)$ term in (\ref{m8.6}) is independent of $a$ and blows up as $R_\theta$ decreases. To see that the force in 6D behaves like $1/a^4$ for very small $R_\theta$ values, consider the Epstein Zeta function $Z_2(R,a/\pi;5)$ term in (\ref{m8.6}). One can expand $Z_2(R,a/\pi;5)$ as, defining $\mu=a/R_\theta$,  
\begin{equation}
 Z_2(1,\sqrt{d};s)=2\zeta(s)+\frac{2\sqrt{\pi}}{d^{(s-1)/2}}\frac{\Gamma(\frac{s-1}{2})}{\Gamma(s/2)}\zeta(s-1) +Q(s,d)
\end{equation}
where 
\begin{equation}
 Q(s,d)=\frac{8\pi^{s/2}}{d^{(s-1)/4}\Gamma(s/2)}\sum_{n=1}^\infty n^{(s-1)/2}\sum_{m=1}^n m^{-s}\sum_{\ell=1}^m \cos\left(\frac{2\pi \ell n}{m}\right) K_{\frac{s-1}{2}}(2\pi \sqrt{d}n).
\end{equation}
In our case, $d=(\mu/\pi)^2$ and $s=5$. Thus, we have
\begin{equation}
 Z_2(1,\mu/\pi;5)=2\zeta(5)+\frac{8\pi^4}{3}\zeta(4)\frac{1}{\mu^4}+Q(5,\mu^2/\pi^2).
\end{equation}
If $R_\theta$ is small for a fixed $a$, we have $\mu\gg 1$ and $Z_2(1,\mu/\pi;5)$ goes to $2\zeta(5)$ which exactly cancels  the other $\zeta(5)$ term in (\ref{m8.6}).  We are left with the $\zeta(4)$ term which is very close to the 4D force term. Note that $Q(s,d)$ also vanishes as $\mu$ increases. For example, the force in 6D becomes, as $\mu\gg 1$
\begin{equation}
 \frac{\mathbb{F}_{6D}^I}{A}\longrightarrow -\frac{\hbar c p \pi^2}{240}\frac{1}{\mu^4 R_\theta^4} +\frac{3\hbar c p}{128 \pi^6}\frac{1}{R_\theta^4}\left[(1+2\mu^2\frac{\partial}{\partial \mu^2})Q(5,\mu^2/\pi^2)\right]
\end{equation}
where the second term vanishes in this limit. This analysis can be generalized to the $q$D case and it can be showed that the same type of cancellations hold. Numerically, for small $R_\theta$ values, we need the precise cancellation of very large numbers which require high numerical precision in the calculation.
\begin{figure}[htb]
\begin{center}
\hspace*{-2.2cm}
 \includegraphics[width=5in,bb=0 282 414 670]{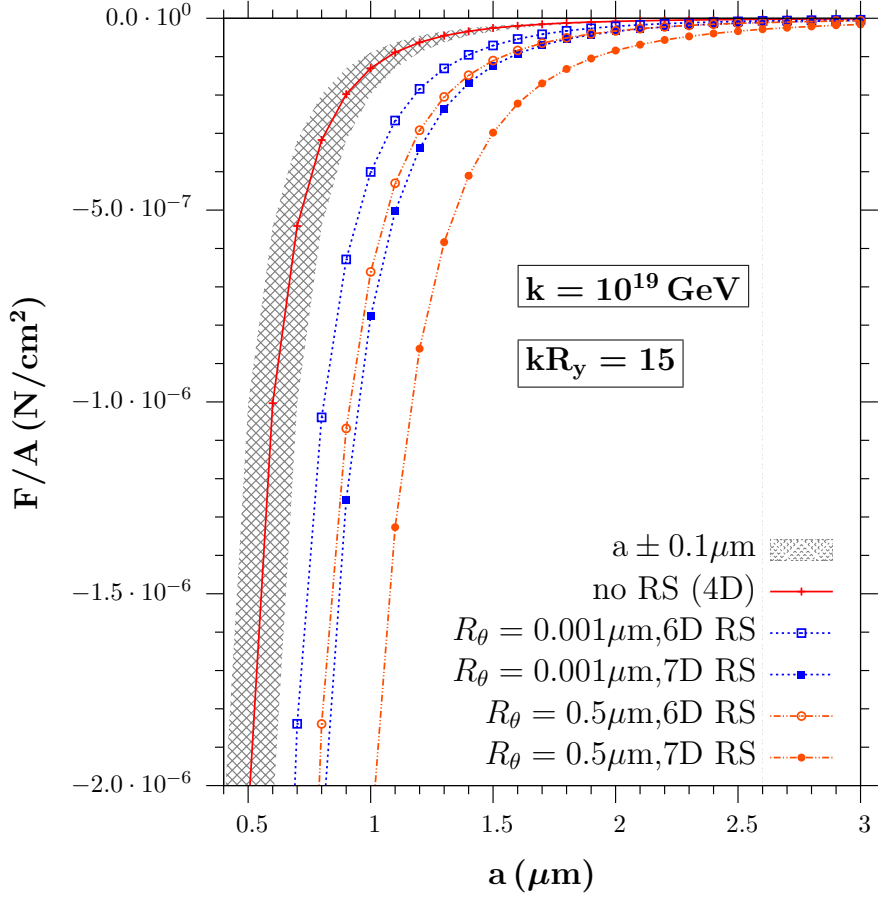}
\end{center}
\vskip -0.2in
\caption{The Casimir force in 6D and 7D as a function of plate separation $a$ for $R_\theta=0.001\mu m$ and $0.5\,\mu m$ with $\mathtt{k}=10^{19}$ GeV and $\mathtt{k}R_y=15$. Here $R_{\theta_1}=R_{\theta_2}=R_{\theta}$ is assumed. The shaded region represents $\pm 0.1\, \mu m$ error for the measurement of plate separation $a$.}\label{7d}
\end{figure}

As can be seen from the above discussion, adding more compact dimensions would not affect much the $\mathtt{k}R_y$ dependence of the force.  Extra dimensions bring an extra suppression factor for the Bessel terms (when $1/R_{\theta_i}$ and $\kappa$ are comparable) but these effects will be minor for most cases. The force will feel the effect of extra $R_\theta$'s in $q$D coming from the terms other than the Bessel functions. As $\mathtt{k}R_y$ is not affeted, in this limit the force is identical to that in UED in $q$ dimensions. There exist some studies in the literature \cite{Poppenhaeger:2003es} as well as some debate on whether the force is attractive or repulsive in extra universal dimensions \cite{Cheng:2006pe}. As shown in previous sections, the 5D form of our force in $\kappa\to 0$ limit is identical to the 5D result of \cite{Poppenhaeger:2003es} in UED and the force is always attractive. In Fig.~\ref{7d} we display the force in both 6D and 7D as a function of $a$ for two values of $R_\theta$, $0.001\,\mu$m and $0.5\,\mu$m. Here we assume $R_{\theta_1}=R_{\theta_2}=R_{\theta}$. We chose $\mathtt{k}R_y=15$,   where there is practically no RS contribution from the Bessel terms so we can maximize the effects from extra compact dimensions. This can be considered like 6D UED in this limit. As in 5D UED, the force remains negative in 6D as well, thanks to the precise cancellation of $a$-independent $1/R_\theta^4$ terms (or $Z_1$ terms in general) with the ones from the $Z_2$ term (similarly between $Z_3$ and $Z_2$ terms as well), as discussed above in 6D RSI. In 5D UED, the bound from Casimir force on $R_\theta$ is very weak (about $0.01\mu$m) \cite{Poppenhaeger:2003es} but adding more compact dimensions, as seen from Fig.~\ref{7d}, will strengthen the bound. Conversely, if the limit on the size of extra dimensions is held constant, their number will be restricted. However, unless one introduces a large number of compact dimensions, it would be unrealistic from Casimir force considerations to get bounds on the size or number of extra dimensions  as strong  as the ones from $b\to s\gamma$, for example \cite{Haisch:2007vb}, which requires $R_\theta$ to be less than $10^{-12}\mu$m in 5D UED.

\section{Summary and Conclusion}
We calculate, using dimensional regularization, the Casimir force between two parallel plates in Randall Sundrum models with extra compact dimensions. We include both the one brane scenario (referred to as  RSII) and the two brane scenario (RSI). We present explicit analytic expressions for the energy between the plates, the renormalized energy, and finally for the Casimir force for the cases of one extra compact dimension (6D), and finally $q$ extra compact dimensions ($q$D) in each model. 

We show consistency of the expressions both from $q$D to 6D and 5D models (RS models with no compact dimensions). We also show that in the limit in which the non-compact dimension of Randall Sundrum models becomes irrelevant, our expressions reproduce exactly the known UED expression in one extra dimension. In that limit, our results present the generalization of the Casimir energy in $q$ universal extra dimensions, and we settle the controversy regarding the sign of the force. We also compare our results with a calculation in RSII models in 6D and $q$D and briefly discuss the discrepancies.

For the RSII model, comparison with the known Casimir force in 4D yields (as in 5D) negligible corrections, making this model not as interesting phenomenologically as RSI. In RSI, we perform an analytical and numerical analysis of our results and discuss the effects of extra compact dimensions on the size of $\mathtt{k}R_y$, the relevant parameter for RSI models, as well as the consequences  of the RS models on the size of the compact (UED like) dimensions. Finally, we analyze the effects of having additional compact dimensions. Although the bounds on the size of the compact dimensions obtained are not as strong as those coming from other low energy experimental restrictions, our results include a comprehensive range of numerical and analytical  explorations of the Casimir force in extra dimensions, and we are consistent with previous calculations in both RS and UED models in the relevant limits.

\section{Acknowledgment}
M.F., I.T. and N.S. are supported in part by NSERC of Canada under the Grants No. SAP01105354 and GP249507. 


\end{document}